\RequirePackage{fix-cm}
\documentclass[twocolumn]{svjour3}          
\smartqed  
\newcommand{\be}{\begin{equation}}
\newcommand{\ee}{\end{equation}}
\newcommand{\ba}{\begin{eqnarray}}
\newcommand{\ea}{\end{eqnarray}}

\newcommand{\beq}{\begin{equation}}
\newcommand{\eeq}{\end{equation}}
\newcommand{\beqn}{\begin{eqnarray}}
\newcommand{\eeqn}{\end{eqnarray}}
\newcommand{\bequ}{\begin{eqnarray*}}
\newcommand{\eequ}{\end{eqnarray*}}
\newcommand{\dt}[1]{ \frac{ d #1}{\ dt}}

\newcommand{\epsfigure}[5]{\begin{figure}%
\centerline{\resizebox{#3}{#4}{\includegraphics{#1/#2.eps}}}%
\caption{#5}%
\label{fig:#2}%
\end{figure}}

\newcommand{\figref}[1]{Fig.~\ref{fig:#1}}

\usepackage{quotes}
\usepackage[dvips]{graphicx}
\usepackage{color}
\usepackage{url}
\usepackage{amsmath}

\usepackage{nomencl}
\makenomenclature

\renewcommand{\dt}[1]{#1'}
\newcommand{\bea}{\begin{equation}\begin{array}{r@{}l}}
\newcommand{\eea}{\end{array}\end{equation}}

\begin{document}

\title{Nonlinear dynamic analysis of an epidemiological model for COVID-19 including public behavior and government action}

\titlerunning{SEIR model with government action}        
\authorrunning{Kwuimy et al.}

\author{C. A. K. Kwuimy \and Foad Nazari \and Xun Jiao \and Pejman Rohani \and C. Nataraj}

\institute{C. A. K. Kwuimy
            \at Department of Engineering Education \\
            University of Cincinnati \\
            \email{cedrick.kwuimy@uc.edu}
         \and Foad Nazari \and Xun Jiao \and C. Nataraj
            \at Villanova Center for Analytics of Dynamic Systems (VCADS) \\
                Villanova University \\
            \email{\{foad.nazari, xun.jiao, nataraj\}@villanova.edu}
         \and Pejman Rohani 
            \at Odum School of Ecology \\ 
            The University of Georgia
}

\date{Received: 12 May 2020 / Accepted: 8 July 2020}

\maketitle

\begin{abstract}
This paper is concerned with nonlinear modeling and analysis of the COVID-19 pandemic currently ravaging the planet.  There are two objectives: to arrive at an appropriate model that captures the collected data faithfully, and to use that as a basis to explore the nonlinear behavior.  We use a nonlinear SEIR (Susceptible, Exposed, Infectious \& Removed) transmission model with added behavioral and government policy dynamics.   We develop a genetic algorithm technique to identify key model parameters employing COVID-19 data from South Korea.  Stability, bifurcations and dynamic behavior are analyzed.  Parametric analysis reveals conditions  for sustained epidemic equilibria to occur.  This work points to the value of nonlinear dynamic analysis in pandemic modeling and  demonstrates the dramatic influence of social and government behavior on disease dynamics.

\keywords{SEIR model \and epidemiology \and COVID-19 \and nonlinear dynamics}
\end{abstract}


\section{Introduction}

Coronavirus disease 2019 (COVID-19) is an infectious disease caused by Severe Acute Respiratory Syndrome CoronaVirus 2 (SARS-CoV-2) that was first identified in China in early December 2019.  It has since become a global pandemic devastating the health, economy and lives of billions of people all over the world and has brought into sharp focus the need for accurate modeling of infectious diseases.  The global government policies are in fact largely being driven by statistical analyses loosely based on nonlinear mathematical models that underlie epidemiology.  As we write this paper, there is also a rising controversy about the predictive power of these models. The crux of the matter is that there is a trade-off between economic disruptions and deaths.  If the model predictions are incorrect in terms of over-prediction, we may be creating mass unemployment and hurting billions of lives by causing economic deprivation.  On the other hand, if the model predictions are wrong through under-prediction, then too many unnecessary deaths would occur. This quandary that most political leaders are finding themselves in points to the need for high accuracy in the models.

Mathematical modeling in epidemiology has a long history dating back to early models by Bernoulli in the eighteenth century \cite{Bernoulli1766,Dietz2002}, although  most current research uses models built on those developed in the 1930s by Kermack and McKendrick \cite{Kermack1927,Kermack1932,Kermack1933}. These are called compartment models, and constitute a set of nonlinear ordinary differential equations, where the state variables represent the population numbers in
various stages of the infectious disease progression, which are described below \cite{Keeling2008}.
\begin{itemize}
\item Susceptible individuals ($S$).  There is no detectable level of pathogens, and the individual's immune system has not developed a specific response to the disease-causing pathogen.
\item Exposed individuals ($E$). The individual has come into contact with an infected person and is infected, but exhibits no obvious symptoms and has low levels of the pathogen that is not high enough to sustain a transmission to other hosts.
\item Infected individuals ($I$).   The number of pathogens has increased to a point that it is now possible to transmit to other susceptible individuals.
\item Removed individuals ($R$).  The individual's immune system has possibly won the battle and reduced the number of parasites significantly and  he/she is no longer infectious.  Or, the individual has been isolated from the population, or, alas, he/she has succumbed to the disease and died.  In all of these cases, the individual is said to be removed.
\end{itemize}

Note that it is common practice to model the number of individuals in each of the above categories as fractions of the nominal population.  We should also observe that other potential variables could be included to account for quarantines, vaccination, etc.
The key factors that govern the dynamics are the growth rate of the pathogen and the level of interaction between the pathogen and the host's immune response.

As in all modeling, we will have to make a compromise between predictive accuracy and complexity.  In addition, since we are using real data, the task of estimating accurate parameters becomes intractable, if not impossible, with a very complex model.  Considering all these factors, we will consider a SEIR model (describing susceptible, exposed, infected and removed individuals) as described further in the sequel.  We will modify the SEIR model with two important features: the effect of government action and that of public reaction.  These two behavioral actions represent social dynamical variables and are especially relevant to the accuracy of predictions as we will show.  Of all the nonlinear phenomena we may expect to find, it is important to note that \emph{endemic equilibrium} points are probably most critical to identify; that is to say, we are interested in knowing under what conditions the disease will \emph{persist}, and not vanish.

Especially with  growing interest in the impact of the COVID-19 virus, there has been an explosion of research papers in modeling and prediction.  It is hence not possible to refer to all - or even a large percentage - of them.  What follows is hence a snapshot focusing somewhat on the subject of the current paper.

As mentioned earlier, the epidemiological models have a rich history after Kermack's original work.  There are several excellent and modern text books \cite{Keeling2008,Brauer2019,Martcheva2015,Vynnycky2010} that describe the fundamental mathematics of epidemiology and discuss the relevance to real historical data of infectious diseases, and we refer the reader to them for a clearer understanding of the model assumptions, derivations and implications.  Hethcote's paper \cite{Hethcote2000} is an especially instructive review and \cite{Heesterbeek2015} is another that skews towards policy decisions.

In terms of nonlinear dynamics, early work by \cite{Grossman1977,Grossman1980} analyzed the effect of seasonal fluctuations as well as contact rate periodicity in what essentially becomes a forced response problem resulting in harmonic and subharmonic resonances.   Several authors have analyzed the occurrence of periodic solutions through Hopf bifurcations in an SEIR model due to the presence of time delays and nonlinear incidence rates \cite{Hethcote1981,Hethcote1991,Bauch2003,Abta2014}.  \cite{Schwartz1983} discovered infinite subharmonic bifurcations in a similar seasonally forced model, while \cite{Yan2014} analyzed bifurcations in the context of limited hospital resources.  \cite{Buonomo2018} is a contemporary review summarizing the literature in seasonal dynamics.
Chaotic motion has also been documented in \cite{Olsen1988,Earn1998}. Finally, \cite{Martcheva2015} provides a clear exposition of nonlinear dynamic phenomena in her monograph.

We should note that the key parameters that can be quite powerful in estimating -- and controlling -- the spread of epidemics, are the so-called reproduction number ($R_0$) and incubation period.  In particular, it can be shown even with the simplest models that the disease will persist if $R_0>1$, and will die out if this number is less than 1.  Much of the control techniques that the governments use are focused on achieving this goal by reducing the transmission rate that eventually controls $R_0$.  From the point of view of mathematical analysis, this creates an interesting situation of a time-varying parameter that is usually discontinuous as government polices are often implemented like step functions.  It should be noted that the incubation period is characteristic of the virus, and is less under our control.  It has been estimated to be 6-7 days \cite{Tang2020,Backer2020}.

As mentioned earlier, COVID-19 has spawned a rich collection of  publications, and we do not deem it necessary to  document them here.  Nevertheless, it is interesting to note the rapid revelations that have come out of these admittedly short-term studies, many of them focusing on data from Wuhan, China, where the virus apparently originated.   WHO \cite{WHO2020} reports that the earliest infections were identified there around the first of December and the infections declined by the end of February  with strong government action as well as public reaction.   The crude fatality rate was estimated to be a shockingly high 3.8\%, although the real number is in all likelihood much lower since the number of infected individuals is heavily undercounted due to logistical limitations in testing, and given that a significant segment of the population is probably infected but asymptomatic.

In quick studies, several researchers \cite{Read2020,LiuT2020,Li2020} have estimated the essential epidemiological parameters using early data from Wuhan, China, in particular, they found $R_0$ to be in the range of 2 to 3.  \cite{LiuY2020} estimated the reproduction number to be 2.7, which is larger than the earlier SARS epidemic, which would make it more dangerous than SARS.  \cite{Kucharski2020} estimated that the travel restrictions that the Chinese government imposed brought down $R_0$ from 2.35 to 1.05, effectively bringing the infections in Wuhan under control.  Even more impressive was the effect of aggressive restrictions on the Diamond Princess cruise ship, which was estimated to reduce $R_0$ from a devastating 14.8 to a more manageable 1.8 \cite{Rockloev2020}.  Several papers have been published attempting to estimate the growth in other areas of China and the world.  \cite{Wu2020} estimated $R_0$ to be 2.7, and predicted similar transmission rates for other cities in China, and \cite{Ferguson2020}, published in mid-March assuming a reproduction number of 2.4, suggested mitigation strategies for various countries, principally US and UK.  This last report was quite influential and led to these two governments to start implementing policies with the objective of ``flattening the curve'' of cumulative infections.

The focus of our study is twofold.
\begin{itemize}
\item We select a data set for COVID-19 that is reasonably complete and accurate and develop a mathematical model that is best able to represent the data.
\item Given the  above fitted model as a starting point, we wish to explore the fundamental nonlinear dynamics of the system and perform a parametric analysis to explore the effect of social dynamics.
\end{itemize}
The reason we use the actual data (in this case, South Korea's) is to keep us grounded in reality and to anchor our parametric studies around this particular situation.  In addition, we expect that a parametric analysis will show the tremendous implication of various actions on the progression of the disease.  In general, our analysis is intended to be relevant to the current situation.  Given that, as of the writing of this paper, the COVID-19 situation is still evolving with considerable uncertainty about the future, we wish to use this paper to validate the importance of mathematical modeling in general, and \emph{nonlinear dynamic analysis in particular}, to enhance our insights.

Building on the above objectives, the rest of the paper is organized as follows.  First, we describe the modified SEIR mathematical model we employ in this study. Next, we describe the data collection and properties. Then, we describe a numerical algorithm we employed and coaxed to get the best parametric fits.  The next section carries out the nonlinear dynamic analysis and describes the interesting results we have achieved.  Finally, we discuss the implications of the model and the results and end with a conclusion.


\section{Notes on Mathematical Models}
\label{sec:model}

We adopt the Susceptible-Exposed-Infectious-Removed (SEIR) framework with a total population size of $N$. In this model, $S$,
$E$, and $I$ represent the susceptible, exposed and infectious
populations and $R$ represents the removed population. For completeness, it is best to start with a standard SEIR model as illustrated in \figref{SEIRmodel1} \cite{Keeling2008,Martcheva2015}.

\bea
\dt{S} & =  \Lambda -\beta{SI} - \mu S \\
\dt{E} & = \beta{SI}-\sigma E  - \mu E  \\
\dt{I} &= \sigma E- \gamma I - \mu I  \\
\dt{R} &=  \gamma I - \mu R
\label{eq:SEIRmodel1}
\eea
where, $'$ denotes derivative with respect to time.

\epsfigure{modelfigures}{SEIRmodel1}{3.0in}{!}{Traditional SEIR Model}


In this model, $\beta$ is the transmission rate, $\mu$ is the death (and emigration) rate, $\sigma$, the incubation rate, is the reciprocal of the latent period (assumed to be the same as the incubation period in this model), and $\gamma$ is the removal rate, and hence the reciprocal of the recovery period (if removal is due to recovery).   Note that $E$ represents those who are exposed but not yet infectious.  We make two modifications to the standard model as described below.

The first modification concerns the specific nature of  COVID-19 and concerns the fact that infected people can be contagious before they show symptoms during the incubation period.  Hence, it is possible that susceptible individuals would have had contact with individuals in both the exposed and infected categories.  
Here, we will model the two  paths from $S$ to $E$ using two values of $\beta$, say $\beta_1$ and $\beta_2$.  Emulating \cite{Gong2003,Hou2020}, we will assume that $\beta_2 = \beta_1 /2$, or that the probability of contacts with asymptotic infected individuals is  half of the  probability of contacts with exposed individuals.  The modified model now becomes:
\bea
\dt{S} & =  \Lambda -\beta_1{SI} -\beta_2 SE - \mu S \\
\dt{E} & = \beta_1{SI} + \beta_2 SE -\sigma E  - \mu E  \\
\dt{I} & = \sigma E- \gamma I - \mu I  \\
\dt{R} & =  \gamma I - \mu R
\label{eq:SEIRmodel2}
\eea

The second modification, illustrated in \figref{SEIRmodel3}, concerns the influence of two important sociological (and  arguably, political) parameters: social behavior and government policy.  Here, we consider the transmission rates to be variable and change with these parameters \cite{Lin2020,He2013}.
Then, the modified model becomes
\bea
\label{eq:SEIRmodel3}
\dt{S} & =  \Lambda - \Upsilon -\mu S \\
\dt{E} & = \Upsilon -\left(\mu+\sigma\right) E \\
\dt{I} & = \sigma E-\left(\mu+\gamma\right)I \\
\dt{R} & =  \gamma I-\mu R  \\
\dt{D} & =  d\gamma I-\lambda D
\eea
where, we have defined an infection function $\Upsilon$ as follows.
\beq \Upsilon=(1-\alpha)\left[\beta_1{SI}(1-D)^{\kappa}+\beta_2{SE}\right]\eeq
Here, $\alpha$ represents the strength of the government action and $\kappa$ is the strength of public response.  Note that  $D$ is a new state variable representing social behavioral dynamics.
$d$ represents the strength of public perception of risk,  $1/\lambda$ is the mean period of public response, and the model reflects the fact that public reaction would increase when more people get infected, and would naturally diminish over time.

\epsfigure{modelfigures}{SEIRmodel3}{3.0in}{!}{SEIR model modified with government action; CI - contact with infected, CE - contact with exposed.}

%


\section{Parameter Identification}
\label{sec:optim}

\subsection{Data}

We use the data from South Korea as our dataset for model fitting for several reasons. Compared to USA, where the testing kits are in significant shortage, and China, in particular Wuhan, where the infected cases went up abruptly in a short period and hence  massive testing might not have been available, the South Korean government was prepared with appropriate emergency measures since January 20th, when it changed its infectious disease alert (in the national crisis management system) category from Level 1 (blue) to Level 2 (yellow)~\cite{CSISpolicy2020}. Such measures provided massive testing capability in South Korea to enable one of the most accurate datasets available.

We examined various databases of South Korea and finally selected COVID-19 Data Repository by the Center for Systems Science and Engineering (CSSE) at Johns Hopkins University~\cite{CSSEdata2020} due to its complete record and accessible interface. In particular, the database provides time series data containing  daily updates on the new infected cases, death cases, and recovered cases, all in a comma-separated values (CSV) file format that is ready to be read and manipulated using standard software tools such as MATLAB.

\subsection{Genetic Algorithm}

Inspired by Charles Darwin's theory of natural evolution, Holland introduced and popularized general-purpose search algorithms that use principles of natural population genetics to evolve solutions to problems, called genetic algorithms (GA) \cite{Holland1975}.
The basic idea in GAs is that evolution will choose the fittest species over  time. Through  emulation of the natural evolution of biological organisms, GA produces a population of individuals (potential solutions in each iteration) to search the solution space of the problem and evolve them through generations to approach the optimal solution. In each generation, the fitness of individuals is evaluated using an objective function and the fittest ones have higher probability to participate in the offspring production process of the next generation. Three main types of operators are employed in GA to guide it towards a solution:
\begin{itemize}
    \item \emph{Selection} to choose between the solutions;
    \item \emph{Mutation} to create and keep genetic diversity; and,
    \item \emph{Crossover} to combine the existing solutions into new ones.
\end{itemize}
Finally, when the stopping criterion is met, the best individual is presented by GA as the solution to the optimization problem.

In this part of the study, the objective is to identify the parameters of the model in such a way that the simulated data matches the real data as much as possible and then use the tuned model to analyze and forecast the spread of COVID-19 in the future. The simulated data is obtained by numerically solving the model in Eq.~\ref{eq:SEIRmodel2} using an integration algorithm (we used sixth order Runge Kutta algorithm).  To accomplish the first part, namely parameter identification, we use GA to find the parameter values which  minimize the cost function between the model prediction and  real data. We devise the cost function based on a weighted sum of the mean square error for both {infected} and {removed} data. Furthermore, as the main purpose of the model is to predict the future, and as the error at the end of the training time span is reflected significantly on the future time evolution, a penalty factor was included in the cost function  for the end points.

The cost function, $f$ is hence defined as follows:
\ba
f & = & W_I  [\mathrm{Mean}(I_r-I_m)^2
+ \alpha_p [(I_r^{t=s} - I_m^{t=s})^2 \nonumber \\
& + & (I_r^{t=e} - I_m^{t=e})^2] + (1 - W_I) [\mathrm{Mean}(R_r - R_m)^2 \nonumber \\
& + & \alpha_p [(R_r^{t=s}-R_m^{t=s})^2 + (R_r^{t=e}-R_m^{t=e})^2] \label{eq:costfunction}
\ea
where $R, I, r, m, s$ and $e$ stand for removed cases, infected cases, real data, model predicted data, start date and end date, respectively. Also, $W_I$ and $1-W_I$ are, respectively, the infected and {removed} patient weights, and $\alpha_p$ is the penalty factor for the end points. For this part of the study, the model described in Section~\ref{sec:model} was used with the assumption that  $\mu$ and $\Lambda$ are zero. The target parameters are $\beta_1$ , $\beta_2$ and $\gamma$ while  $\sigma$ was assumed to be 0.14, equivalent to incubation period of 7 days for COVID-19 \cite{Tang2020a}.  We will consider the population to be constant; in other words, we assume that $N$ does not change.  This means that the natural mortality (including emigration/immigration) rate ($\mu$) as well as the birth rate ($\Lambda$) are zero. This assumption is reasonable over the short time period of analysis.

As explained earlier in this study,  factors like government actions can significantly affect the trend and pattern of disease spread and accordingly, the SEIR model parameters. So, in this study we solved the  parameter identification problem for two separate time spans, i.e., controlled and uncontrolled \cite{DeCastro2020}.

Totally, the recorded data of 108 days was employed in this study (Jan 22, 2020 to May 8, 2020). The uncontrolled data was taken for first 40 days (Jan 22, 2020 to March 1, 2020) and controlled data for the next 68 days (March 2, 2020 to May 8, 2020). Out of 68 days of controlled time span, first 40 days were used for model tuning and the next 28 days for evaluating the performance of the model in forecasting the unseen data. The total population of South Korea was taken to be 51,269,185 from standard sources.

\subsection{Results}

\begin{figure}[tp]
\begin{center}
\includegraphics[scale=0.5]{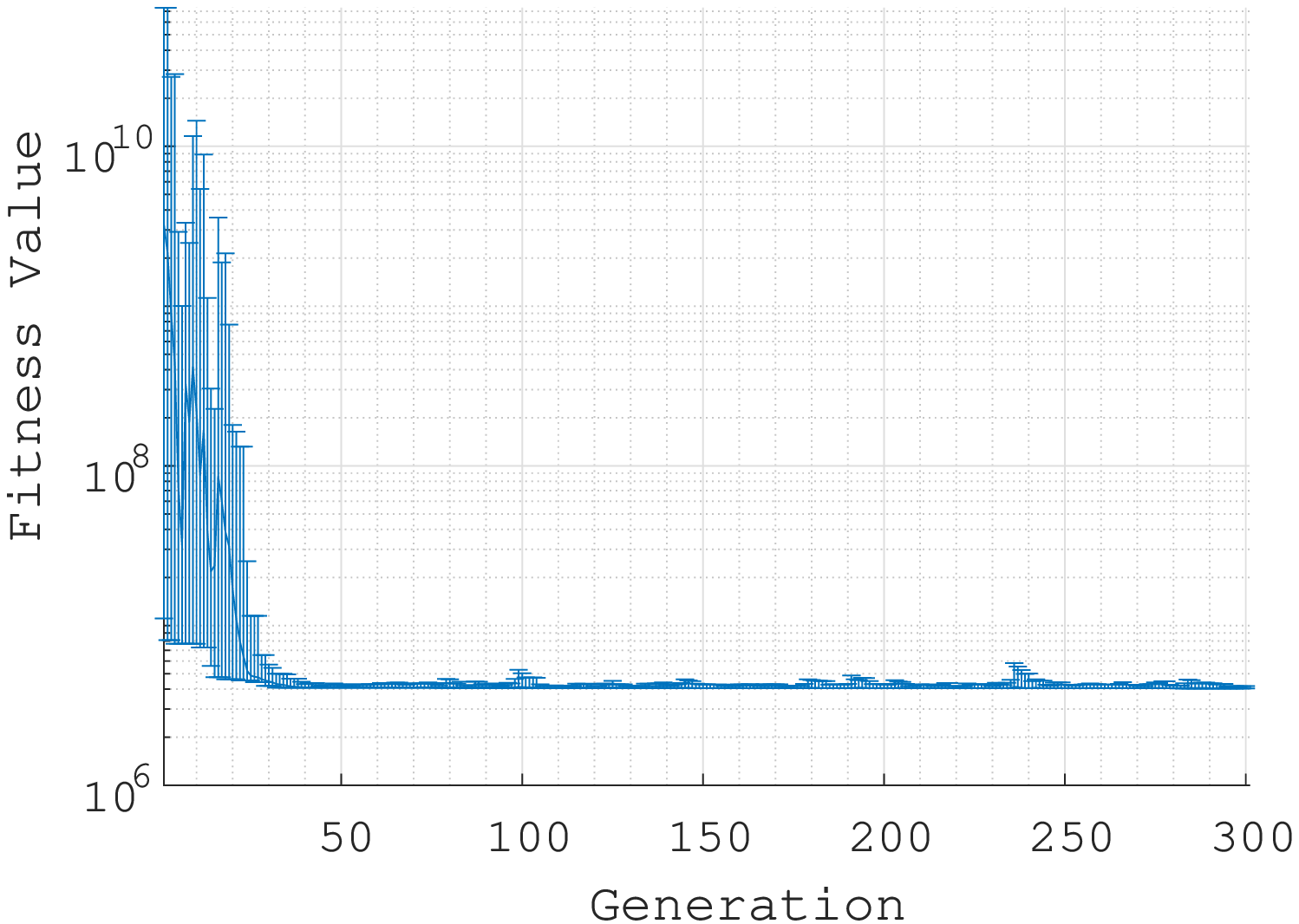}
\end{center}
\caption{GA optimization convergence over the controlled time span of COVID-19 spread in South Korea}
\label{GATrend}
\end{figure}

The process of optimum selection of the optimization variables was accomplished with a population size of 200 with a cross-over probability of 0.8 for 300 generations. For uncontrolled and controlled time spans, the value $\alpha_p$ and $W_I$ was 10 and 0.5, respectively. $W_I=0.5$ means that the model predicted removed and infection rates have the same weights in the cost function and so the optimization algorithm tries to make both of them close to the real data, simultaneously and equally. Also,  $\alpha_p=10$ means that the square error of model infection and removed rates at each end point has 10 times more effect on the cost function than the mean square error of all 40 days and this enforces the model to be close to the real data at the end points. The initial number of infected and removed cases for the SEIR model in both periods was considered as the real data, i.e., $I_0=1$ and $R_0=0$ for uncontrolled and $I_0=3736$ and $R_0=47$ for controlled time span. Also, due to lack of $E_0$ (initial number of exposed individuals) in the available dataset, it was assumed to be two times  $I_0$. The trend of optimal tuning of the model parameters for the controlled time spans is shown in Figure~\ref{GATrend}. The convergence of the best fitness value range, including best, mean and worst values, to an optimum condition over 300 generations is demonstrated in the logarithmic form in this figure.

\begin{figure}[tp]
\begin{center}
\includegraphics[scale=0.5]{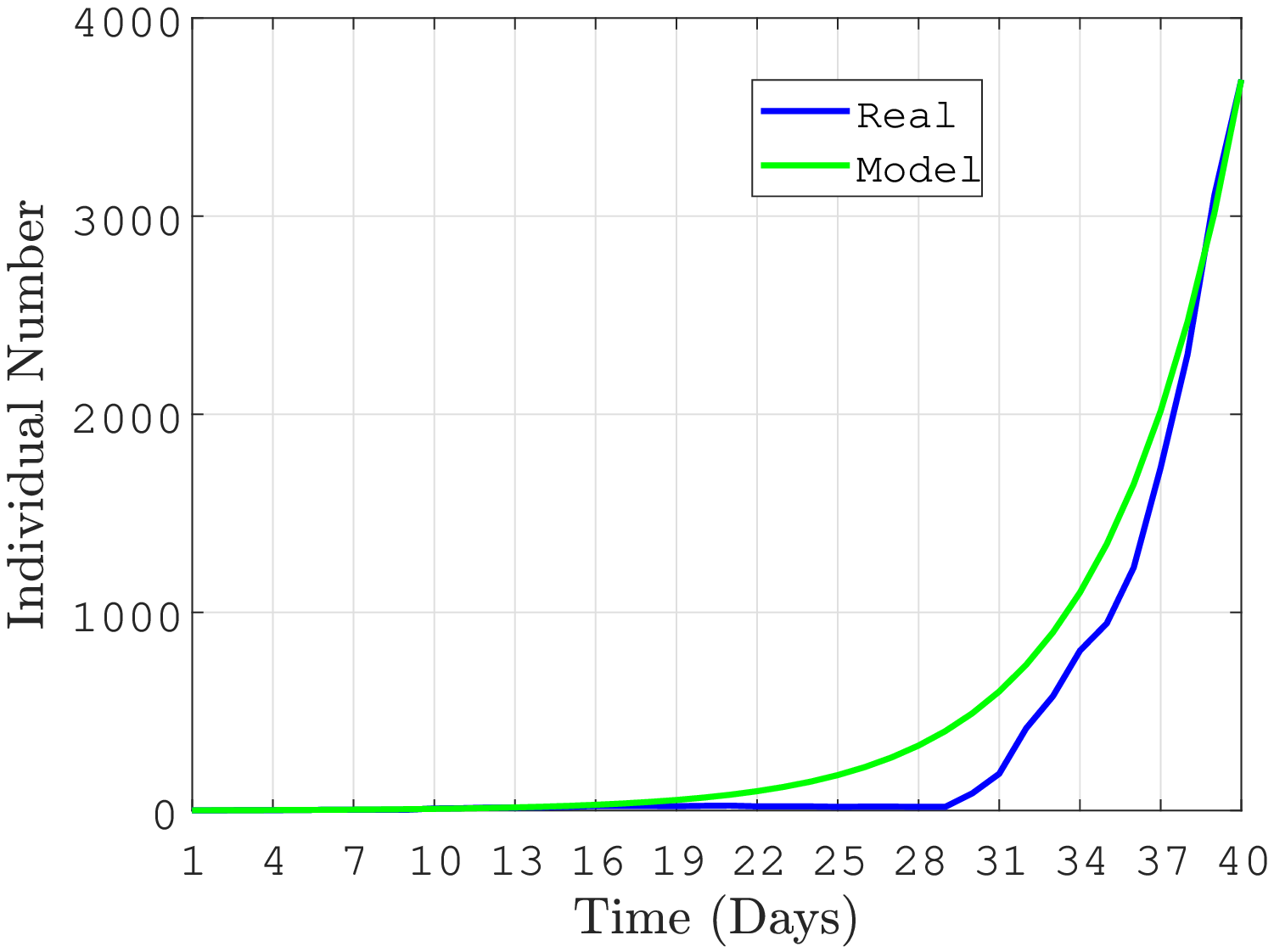}
(a)
\includegraphics[scale=0.5]{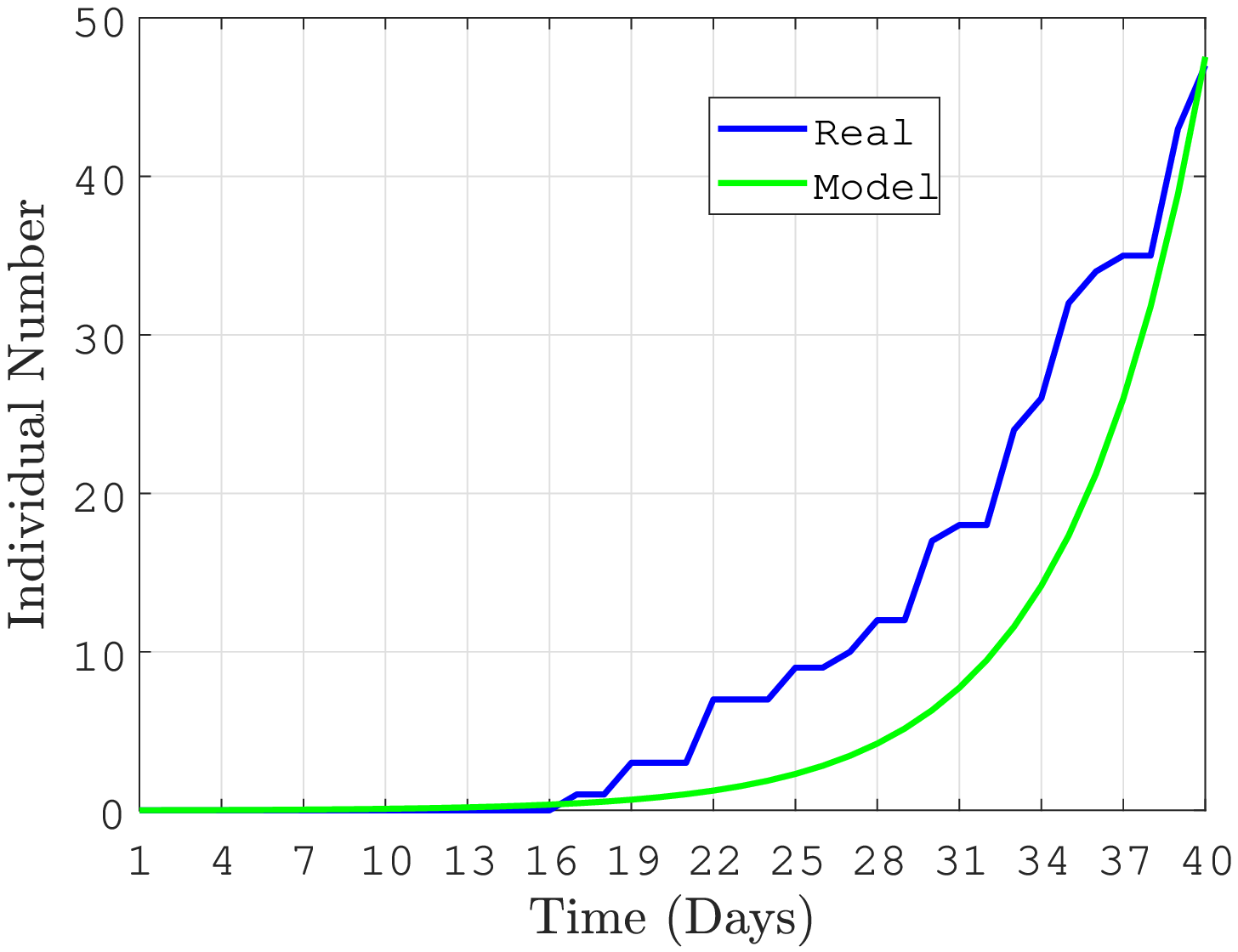}
(b)
\end{center}
\caption{Real and simulated (a): infected and (b): removed individuals for the \emph{uncontrolled} time span of COVID-19 spread in South Korea}
\label{ModelvsReal_U}
\end{figure}

\begin{figure}[tp]
\begin{center}
\includegraphics[scale=0.5]{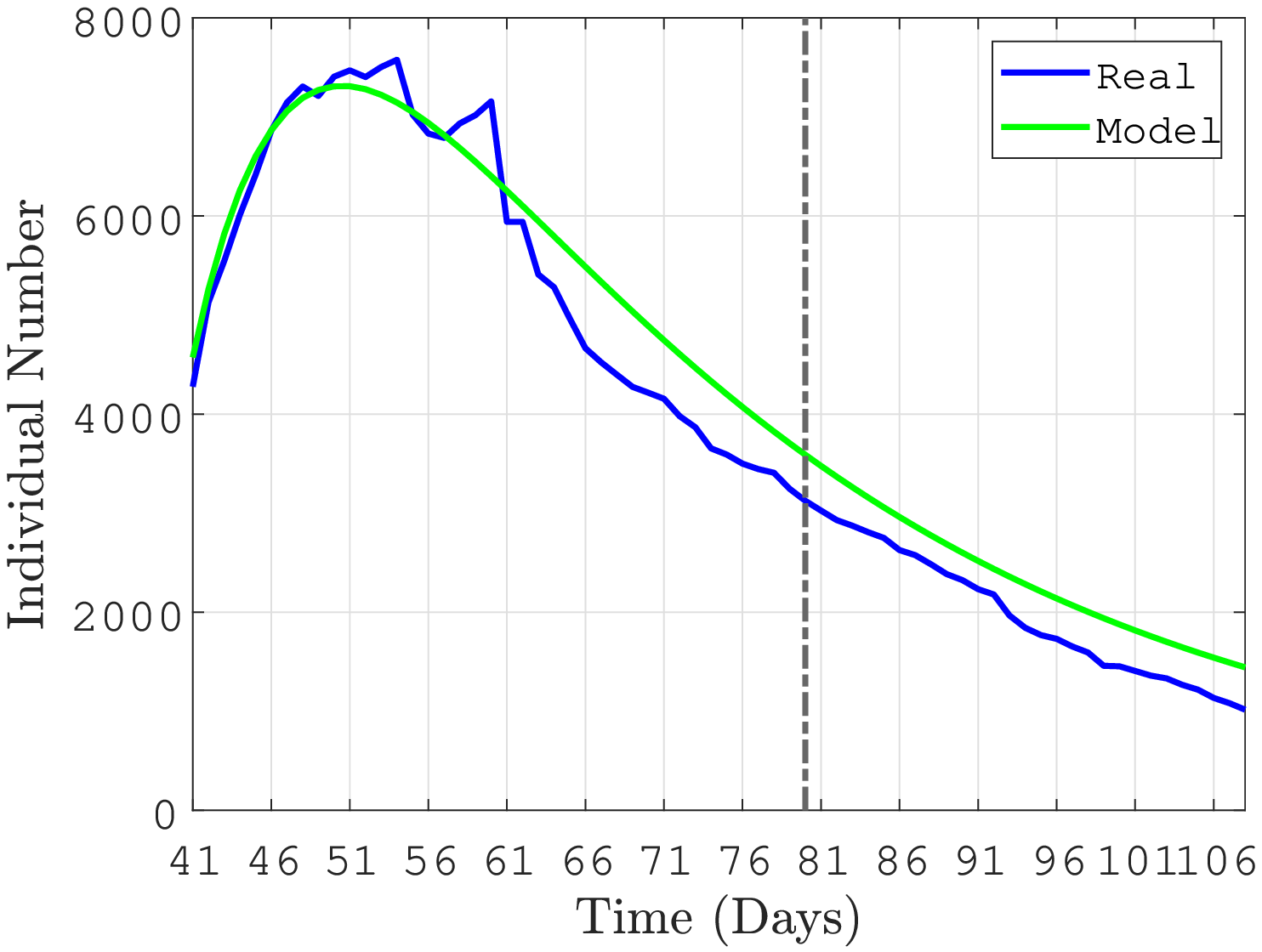}
(a)
\includegraphics[scale=0.5]{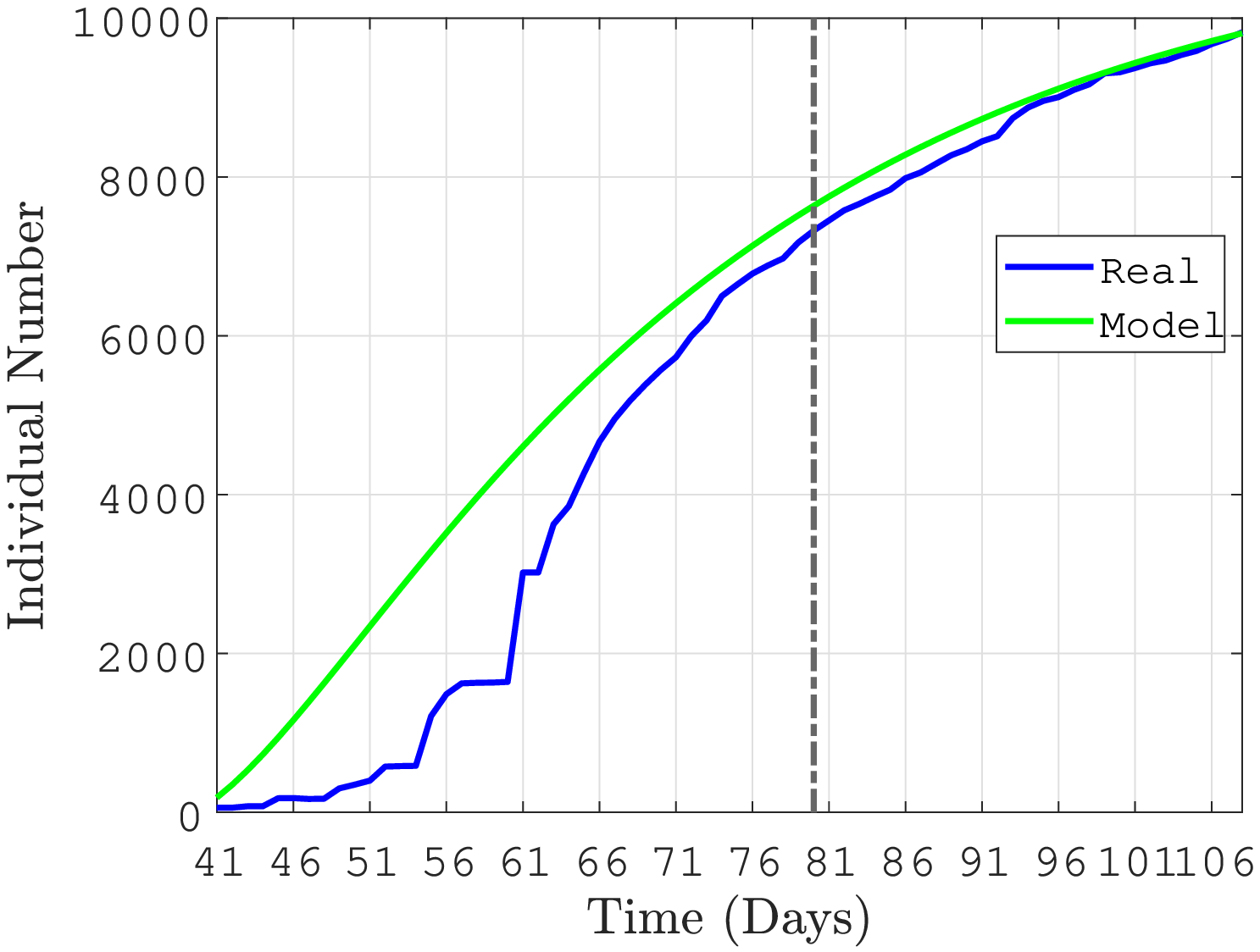}
(b)
\end{center}
\caption{Real and simulated (a): infected and (b): removed individuals for the \emph{controlled} time span of COVID-19 spread in South Korea}
\label{ModelvsReal_C}
\end{figure}

The GA parameter identification results are tabulated in  Table~\ref{GAresults}. As can be seen in this table, the values of model parameters  changed significantly with transition from uncontrolled to controlled time span owing to strong actions which were imposed to control the disease transmission in South Korea. The effect of this change in the parameter values is clearly seen in the comparison between the trend of individual numbers in the uncontrolled (Figure~\ref{ModelvsReal_U}) and controlled (Figure~\ref{ModelvsReal_C}) time spans. The sharp drop in the number of active infected individuals and reduction in the growing slope of accumulative removed cases shows that the actions that have been taken in this country to control the COVID-19 spread have been quite successful.

\begin{table*}[tbp]
\caption{GA parameter identification results for uncontrolled and controlled time spans}
\begin{center}
\begin{tabular}{ | c | c | c | c | }
\hline
Condition& $\beta_1$ & $\beta_2$ & $\gamma$ \\
\hline
Uncontrolled & 0.4071399 & 0.0626798 & 0.0026 \\
\hline
Controlled & 9.98E-07 & 7.66E-06 & 0.0329 \\
\hline
\end{tabular}
\label{GAresults}
\end{center}
\end{table*}

Comparison of model and real data for the first 80 days indicates that parameter identification for both uncontrolled and controlled conditions has been performed, appropriately, and there is good agreement between them. Also, it is observed that the model was able to forecast the unseen data of days 81 to 108 quite well. Nevertheless,  there is still a difference between real and model predicted data.  Some possible reasons include perhaps overly simplistic modeling of sociological behavior and government actions and inaccuracy in the assumed model parameters like $\sigma$, $E_0$, $\Lambda$ and $\mu$. It should also be noted that the number of infected individuals is a measure of the amount of testing that was done, which has not been  comprehensive, and hence the numbers can be inaccurate.  For all these reasons, the fluctuations seen in the real data are not predicted precisely by the model, but it is clear that the general trends of variation of both infection and removed rates are quite similar.

%
%
%
%

\section{Nonlinear analysis}

This section focuses on the disease extinction or persistence, which is determined by the stability of the disease free equilibrium and the existence of endemic equilibrium. Prevention and control of COVID‐19
epidemics require a better understanding of its mode of dissemination as well as the impacts of control strategies. The analysis considers a na\"{i}ve scenario where there is no governmental action, which is unlikely but will provide a baseline to appreciate the effects of the action. In the second and third scenarios we consider the effects of individual reaction and the governmental action.

\subsection{System without controlling action}

In this scenario, the infection function is given by
\beq\label{eq:SEIR41} \Upsilon=\left[\beta_1{SI}+\beta_2{SE}\right]\eeq
which captures the possibilities of new infection by both infected and exposed individuals. The corresponding model is shown in Eq.~\ref{eq:SEIRmodel2}.

\textbf{Proposition 1:} The disease free equilibrium $E^0=(\Lambda/\mu,0,0,0)$ of model Eq.~(\ref{eq:SEIRmodel2}) is asymptotically stable if $R_0<1$. The endemic equilibrium $E^1=(S_0,E_0,I_0,R_0)$ is asymptotically stable if $R_0>1$, and is defined as:
\beqn\label{eq:SEIR4}
E_0^*&=&\frac{\mu(\mu+\gamma)}{\beta_1\sigma+\beta_2(\gamma+\mu)}(R_0-1)\nonumber\\
R_0^*&=&\frac{\gamma\sigma}{\beta_1\sigma+\beta_2(\gamma+\mu)}(R_0-1);\nonumber\\
S_0^*&=&\frac{(\sigma+\mu)(\gamma+\mu)}{\beta_1\sigma+\beta_2(\gamma+\mu)}; \nonumber\\ I_0^*&=&\frac{\sigma\mu}{\beta_1\sigma+\beta_2(\gamma+\mu)}(R_0-1)
\eeqn
and
\beq \label{eq:SEIR53}
R_0=\frac{\Lambda\left[\beta_1\sigma+(\gamma+\mu)\beta_2\right]}{(\mu+\sigma)(\mu+\gamma)\mu}\eeq

\textbf{Proof:}
In order to compute the expression of the equilibrium points, we set the time derivative to zero (steady state) and solve the corresponding algebraic equation.
\beqn\label{eq:SEIR3}
0 & = & \Lambda -\beta_1(1-\alpha){SI}-\beta_2(1-\alpha){SE}-\mu S \nonumber \\
0 & = &\beta_1(1-\alpha){SI}+\beta_2(1-\alpha){SE}-\left(\mu+\sigma\right) E \nonumber \\
0 & = &\sigma E-\left(\mu+\gamma\right)I \nonumber \\
0 & = & \gamma I-\mu R
\eeqn
It is obvious that $E^0(\Lambda/\mu,0,0,0)$ is a trivial solution of Eq.~(\ref{eq:SEIR3}). $E^0$ is called the disease free equilibrium since it is obtained for $I=E=0$ and the corresponding infected function $\Upsilon$ is zero. For $I\neq0$, the model in Eq.~(\ref{eq:SEIRmodel2}) has a non-zero solution $E^1$ given by Eqs.~(\ref{eq:SEIR4}).

The stability of the equilibrium points $E^0$ and $E^1$ is obtained from the Routh–Hurwitz criterion for stability, which states that the equilibrium state is stable if the roots of the characteristic polynomial in $\zeta$ are all negative. The Jacobian matrix of the system is obtained as
\beq\label{eq:SEIR51} J=\left(
         \begin{array}{cccc}
           -a_{11}-\mu & -a_{12} & -a_{13}&0  \nonumber\\
           a_{11} & a_{12}-(\mu+\sigma) & a_{13} &0 \nonumber\\
         0 & \sigma & -a_{33}&0\\
           0&0&\gamma&-\mu\nonumber\\
\end{array}
       \right)\eeq
with
\beqn\label{eq:SEIR6}
a_{11} &=& \beta_1S_0+\beta_2S_0;\nonumber\\
a_{12} &=& \beta_2S_0;\nonumber\\
a_{13} &=& \beta_1S_0;\nonumber\\
a_{33}&=&\gamma+\mu.
\eeqn
The characteristic polynomial for the DFE is the following ($E_0=I_0=0$ and $S_0=\Lambda/\mu$)
\beq\label{eq:SEIR7} (\zeta+\mu)\left[\zeta^2+a_1\zeta+a_0\right]\eeq
with
\beqn\label{eq:SEIR8}
a_1 &=& -\gamma-\sigma-2\mu+\beta_2S_0\nonumber\\
a_0&=&\left[\beta_1\sigma+\beta_2(\gamma+\mu)\right]\frac{\Lambda}{\mu}\frac{1-R_0}{R_0}\eeqn

The system is stable if the roots of the characteristic equation Eq.~(\ref{eq:SEIR7}) are all negative; this is satisfied if $R_0<1$, which is equivalent to
  \beqn
\beta_1\sigma+\beta_2(\gamma+\mu)<\frac{\mu}{\Lambda}(\gamma+\mu)(\sigma+\mu)\eeqn
For the endemic equilibrium, the steady state system in Eq.~(\ref{eq:SEIR3}) can be solved to obtain Eq.~({\ref{eq:SEIR4}). The coefficients of the characteristic polynomial are given as
\beqn
a_2&=&\beta_2S_0-\mu(R_0-1)-(\gamma+\sigma+3\mu)\nonumber\\
a_1&=&a_{33}(a_2+a_{33})+\sigma a_{13}+\frac{a_0-a_{13}\sigma\mu}{a_{33}}\nonumber\\
a_0&=&\mu(\gamma+\mu)(\sigma+\mu)(R_0-1)
\eeqn

The system is stable is the roots of the characteristic polynomial are all negative; that is if $R_0>1$, which is equivalent to
\beqn
\beta_1\sigma+\beta_2(\gamma+\mu)>\frac{\mu}{\Lambda}(\gamma+\mu)(\sigma+\mu)\eeqn

Figure~\ref{fig1}a shows an illustration of a DFE situation where $R_0=0.7$ and $\beta_2=0.0517$, $\beta_1=0.0024$, $\sigma=0.14$ and $\gamma=0.0026$. Using the transmission rate coefficients obtained from Section~\ref{sec:optim} ($\beta_2=0.0628$, $\beta_1=0.407$) we get the endemic equilibrium of Fig.~\ref{fig1}b. The effects of the transmission rates $\beta_1$ and $\beta_2$ are illustrated in Fig.~\ref{fig1}b. The figure considers the situation of fewer contacts with infected individuals ($\beta_1<\beta_2$, most/some infected individual are in quarantine assuming the same probability of contamination once in close contact), and compares it to the situation where we have higher probability of contamination with infected individuals, or $\beta_1>\beta_2$). Beyond $R_0=1$, the proportion of infected individuals naturally increases and is higher when $\beta_2>\beta_1$. This can be interpreted to mean that exposed people will have a greater impact on the persistence of the disease. These results confirm the observations of the number of newly confirmed cases due to close contact with exposed and infected individuals in Wuhan, China \cite{Hou2020}.

\begin{figure}[!ht]
    \begin{center}
        \includegraphics[scale=0.35]{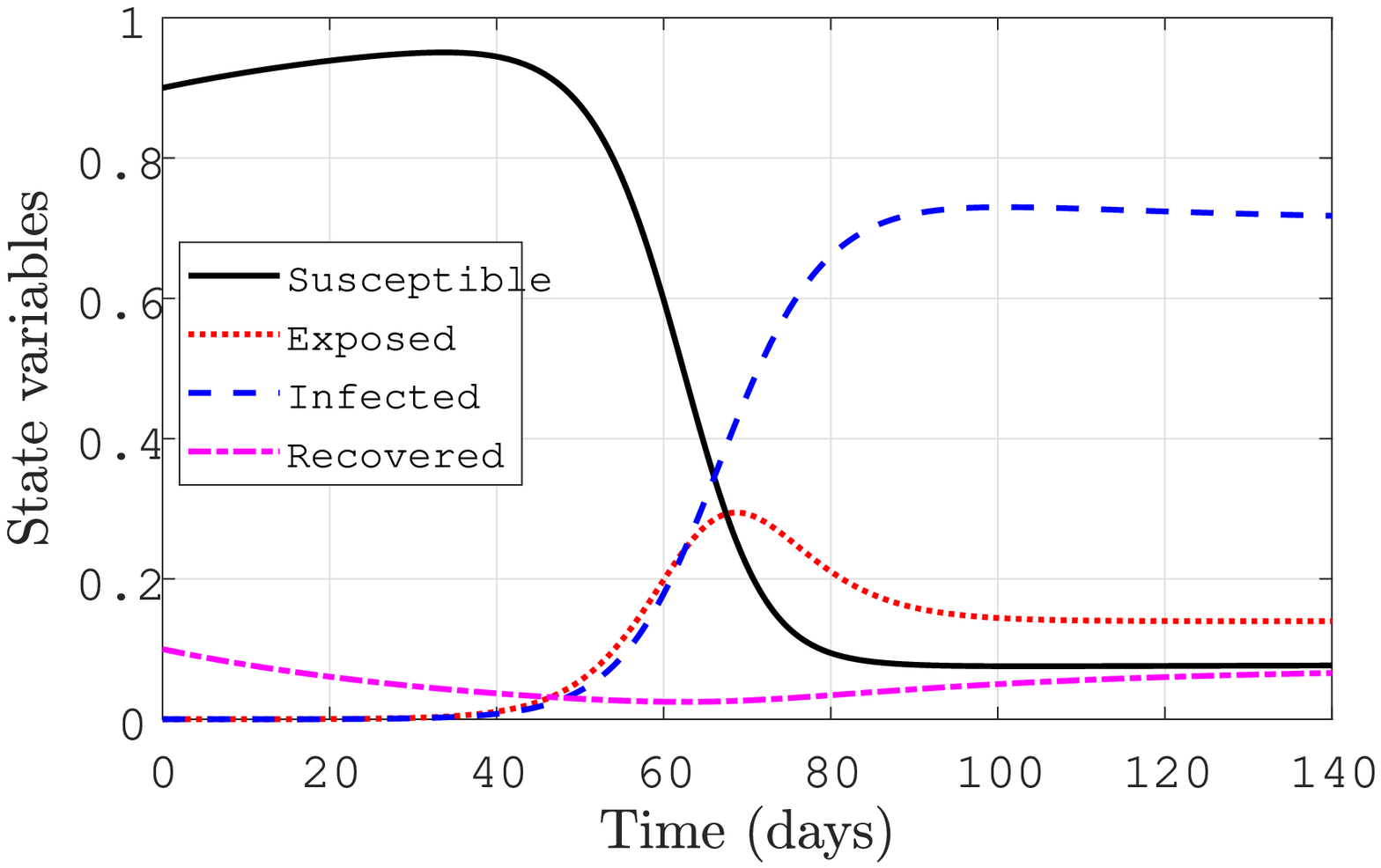}(a)\\
           \includegraphics[scale=0.40]{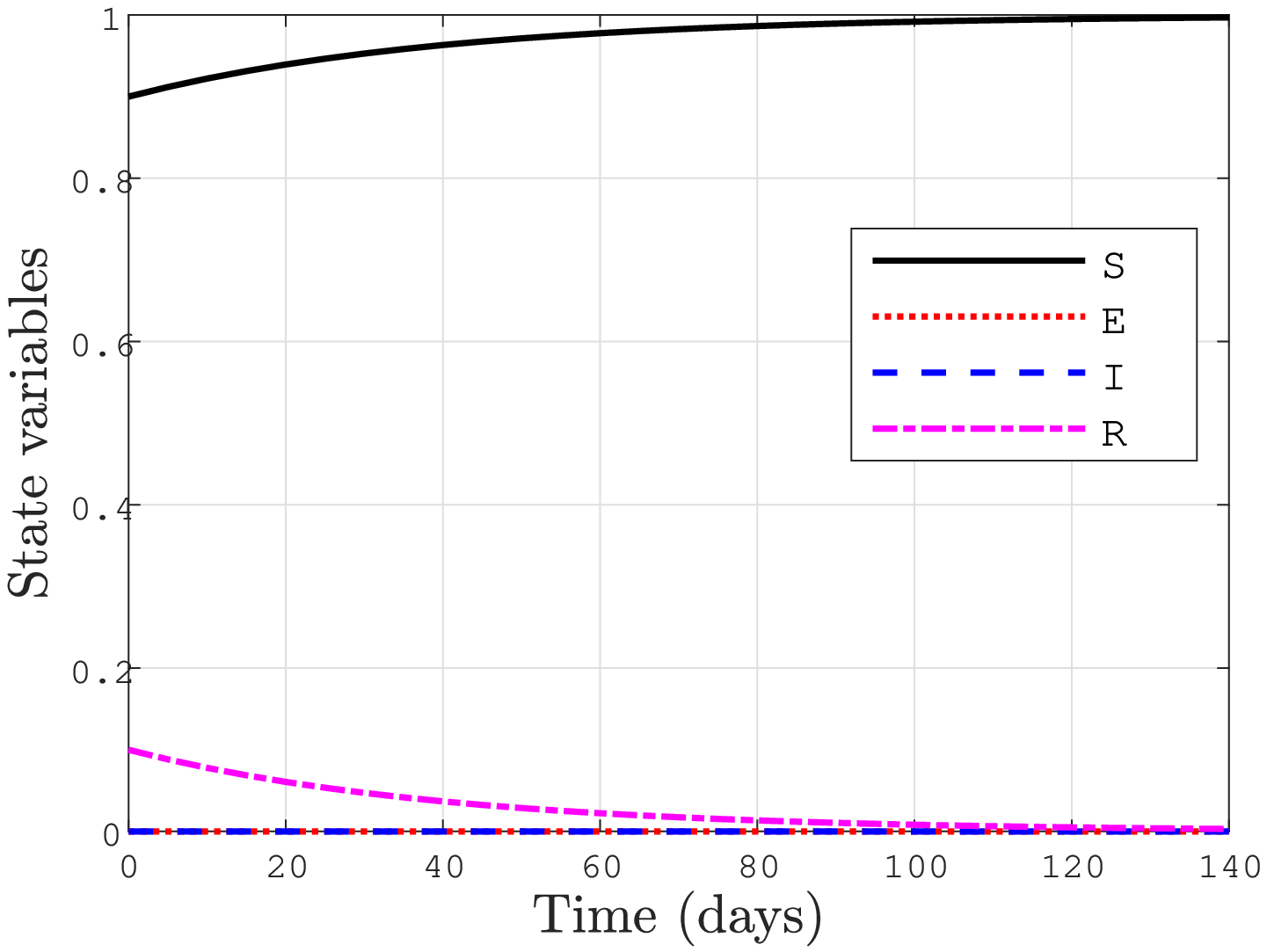}(b)\\
\caption{Response of the system for $R_0=0.7$. (a) DFE $\beta_2=0.0517$, $\beta_1=0.0024$, (b) Endemic equilibrium $\beta_2=0.0628$, $\beta_1=0.407$.}
\label{fig1}\end{center}
   \end{figure}

\begin{figure}[!ht]
    \begin{center}
        \includegraphics[scale=0.35]{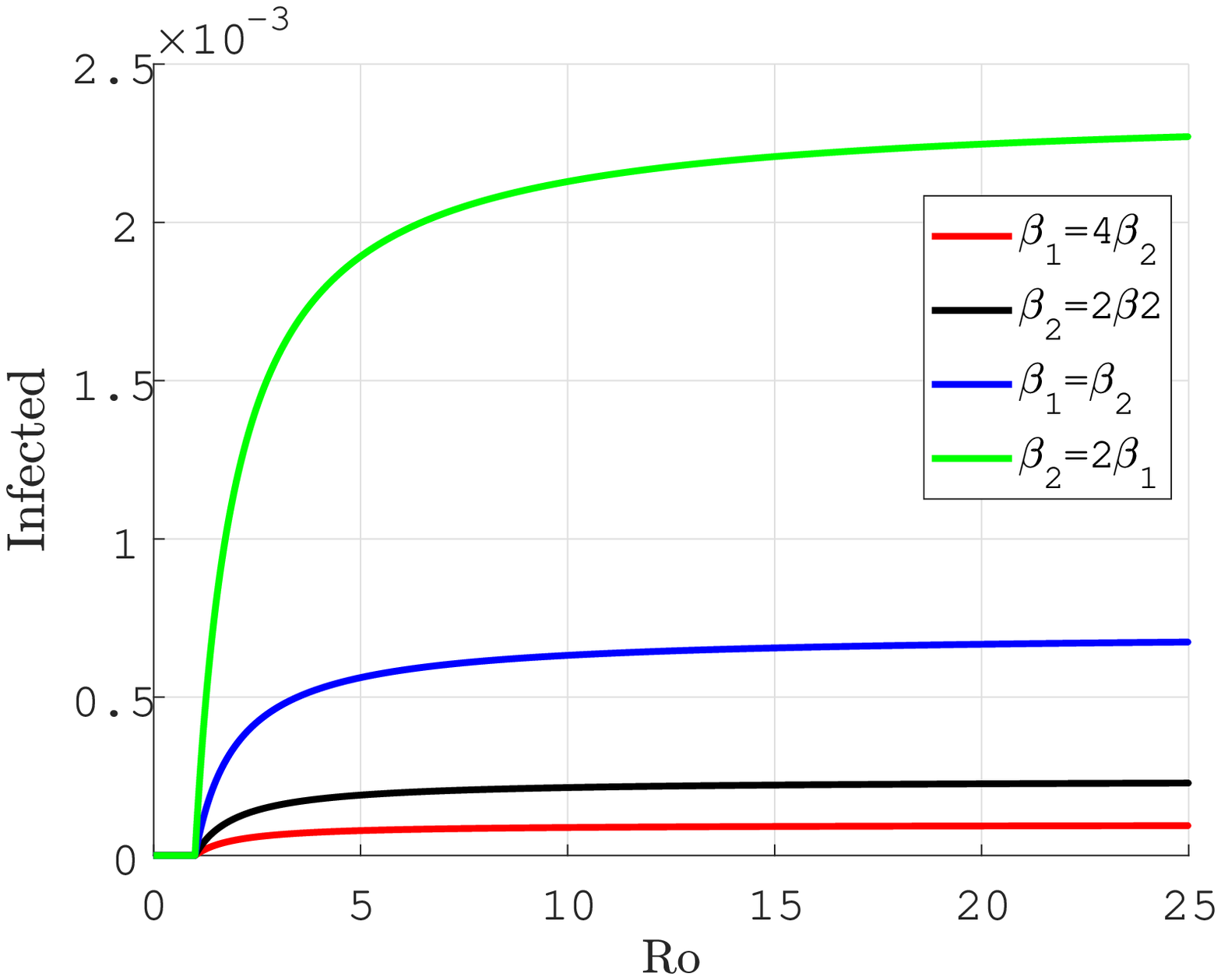}
\caption{Stable DFE and endemic equilibrium as function of the transmission rates $\beta_1$ and $\beta_2$.}
\label{fig2}\end{center}
   \end{figure}

\subsection{Effects of governmental action}

In the context of COVID-19, governmental actions are mainly focussed on regulating social life to reduce the likelihood of contact between individuals. This naturally impacts the transmission rates. The effects of governmental actions are summarized in the infection function, which would need to be substituted into Eq.~\ref{eq:SEIRmodel3}; however, note that we do not consider the effect of public reaction here, hence we drop $D$ from the equation.
\beq \Upsilon=(1-\alpha)\left[\beta_1{SI}+\beta_2{SE}\right]\eeq

\textbf{Proposition 2:}
In this case, an endemic equilibrium ($R_0>1$) is persistent if
\beq0<\alpha<\alpha_c=1-\frac{1}{R_0}\eeq

\textbf{Proof:}
Under the effect of governmental action, repeating the analysis in the previous paragraph will not change the DFE, endemic equilibrium and the stability conditions if $\beta_i$ is replaced by $(1-\alpha)\beta_i$, ($i=1,2$). However, the reproduction number becomes
\beq \label{eq:SEIR52}
R_{00}=(1-\alpha)R_0=\frac{\Lambda(1-\alpha)\left[\beta_1\sigma+(\gamma+\mu)\beta_2\right]}{(\mu+\sigma)(\mu+\gamma)\mu}\eeq
The endemic equilibrium is stable if $R_{00}>1$, which leads to the critical value of the governmental control
\beq0<\alpha<\alpha_c=1-\frac{1}{R_0}\eeq
\begin{figure}[!ht]
    \begin{center}
        \includegraphics[scale=0.35]{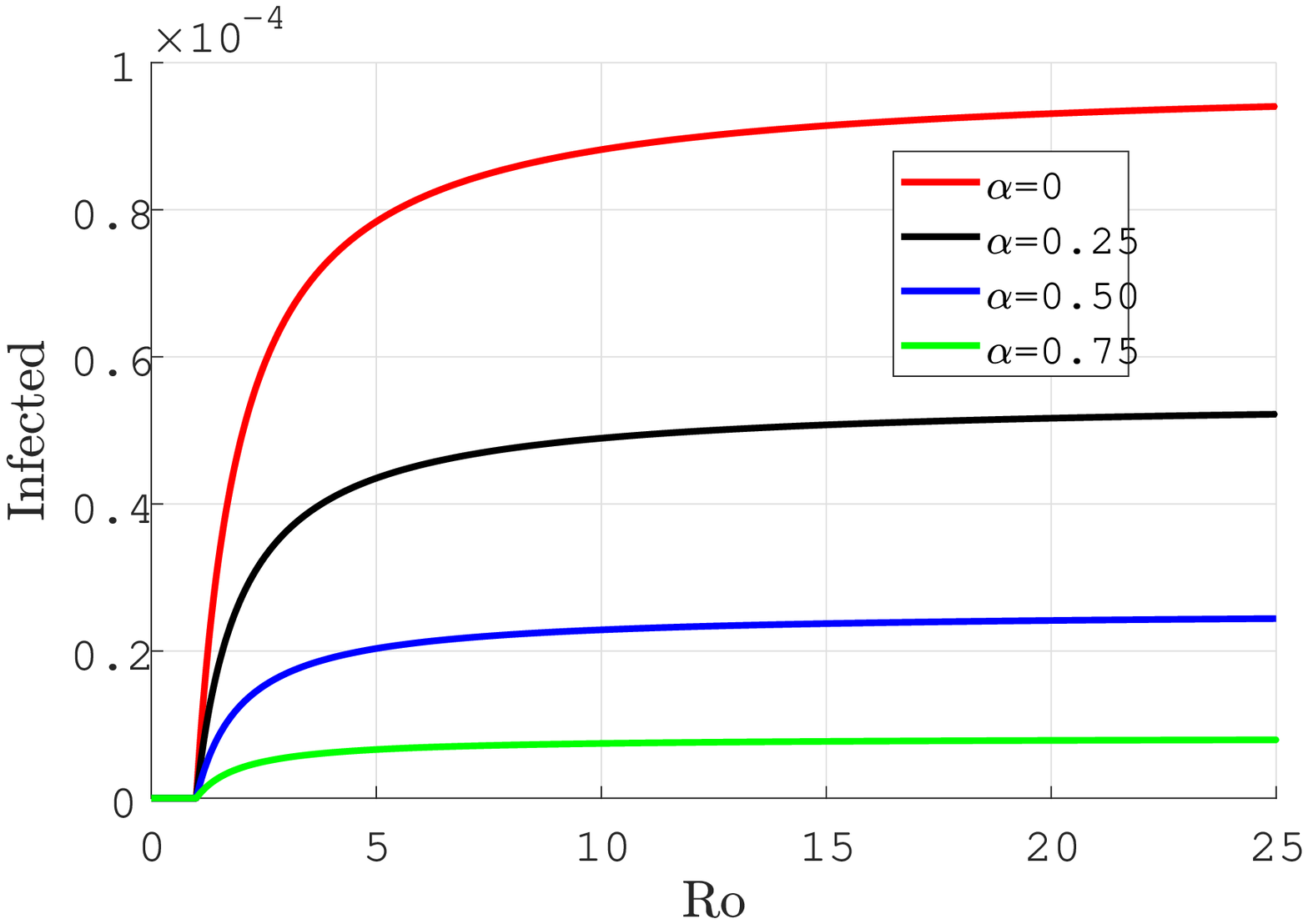}\\(a)
            \includegraphics[scale=0.35]{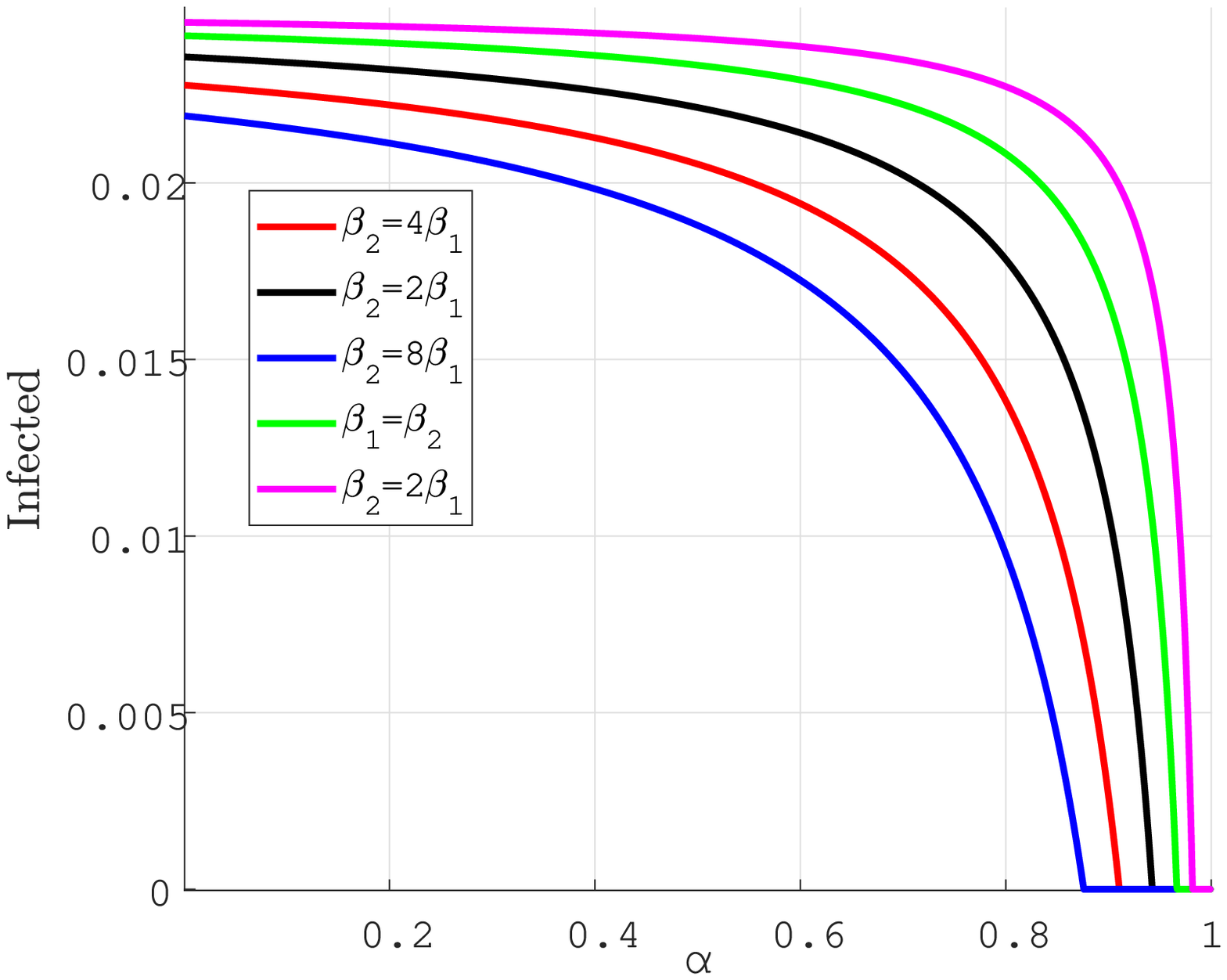}\\(b)
\caption{Effects of the governmental control on the endemic equilibrium for the values of Fig~\ref{fig1}. (a): Effects of $\alpha$ on the Reproduction number for $\beta_1=4\beta_2$. (b): Effects of $\alpha$ on the proportion of infected individuals for $R_0>1$ compute from the equation.}
\label{fig3}\end{center}
   \end{figure}

Figure~\ref{fig3} gives two different views of how the government action could contribute to control the spread of the disease. As might be expected, stronger governmental action (higher values of $\alpha$) has more impact on the disease (Fig.~\ref{fig3}a). But, what is more interesting is that the results predict the existence of a threshold value $\alpha_c$ expressed as a function of the transmission rate, that would lead to complete control of the disease. This threshold value is higher for lager values of $\beta_2$. In practice of course, there would be a natural limit to the governmental action. For this reason, additional controls would be needed. The literature that documents past infectious diseases similar to the COVID-19 have shown how an increase in the number of deaths and the severity of critical cases can be leveraged to impact the perception and seriousness of the population.

\subsection{System with additional control}

We now take into consideration the combined effects of the government action and the public perception of risk regarding the number of severe and critical cases. The variable $D$ is added to the model to represent the public perception of risk. It increases when people die, and will decay naturally, meaning that perception of risk diminishes over time in the absence of the COVID-19. The intensity of this perception is carried through the intensity of the population response $\kappa$ and proportion of severe cases $d$. The infection function is  now
\beq \Upsilon=(1-\alpha)\left[\beta_1{SI}(1-D)^{\kappa}+\beta_2{SE}\right]\eeq
which would be substituted into the model given by Eq.~\ref{eq:SEIRmodel3}.


\textbf{Proposition 3: }
The system in Eq.~\ref{eq:SEIRmodel3} under control has a higher threshold for the onset of endemic equilibrium. This onset value is $R_0=1$ for $\kappa=0$ and $R_0>1+\kappa_{\mathrm{effect}}$ for large values of $\kappa$, where
\beq \kappa_{\mathrm{effect}}=\frac{\beta_1\sigma}{\beta_2(\gamma+\mu)}\eeq

\textbf{Proposition 4: }
There is an endemic state for which the intensity of the public perception has no effect. That endemic state is defined by:
\beqn
I_{c0}^* & = & \frac{\lambda}{d\gamma}<1, \quad \rightarrow~~E_{c0}^*=\frac{\gamma+\mu}{\sigma}\frac{\lambda}{d\gamma},  \nonumber \\ R_{c0}^*&=&\frac{\gamma}{\mu}\frac{\lambda}{d\gamma}, \nonumber \\
S_{c0} & = & \frac{\Lambda}{R_0\mu}\left[1+\frac{\beta_1\sigma}{\beta_2(\gamma+\mu)}\right]
\eeqn

\textbf{Proof:}
The steady state conditions lead to
\beqn\label{eq:SEIR42}
E_{c0}^*&=&\frac{\mu+\gamma}{\sigma}I_{c0}^*; \quad
R_{c0}^*=\frac{\gamma}{\sigma}I_{c0}^*; \quad 
D_{c0}^*=\frac{d\gamma}{\lambda}I_{c0}^*;\nonumber\\
S_{c0}^*&=&\frac{\Lambda}{R_0\mu}\frac{\beta_1\sigma+\beta_2(\gamma+\mu)} {\beta_1\sigma\left(1-\frac{d\gamma}{\lambda}I_{c0}\right)^{\kappa}+\beta_2(\gamma+\mu)}; \nonumber\\
I_{c0}^*&=&\frac{I_0^*}{R_0-1}\left[R_0  \right. \nonumber \\
 & &   \left. \quad \quad -\frac{\Lambda}{R_0\mu}\frac{\beta_1\sigma+ \beta_2(\gamma+\mu)}{\beta_1\sigma\left(1-\frac{d\gamma}{\lambda}I_{c0}\right)^{\kappa}+ \beta_2(\gamma+\mu)}\right]
\eeqn
where, the subscript $c$ stands for control.  

The transcendental equation would not lead to an explicit expression of $I_{c0}$. Thus, guided by the literature, we limit the analysis to some specific cases.
\bea
\textrm{For}~~\kappa=0, \rightarrow \quad & I_{c0}^*=I_0,~~E_{c0}^*=E_0, \\
                                          & R_{c0}^*=R_0, ~~S_{c0}^*=S_0
\eea
For $ \kappa~~\rightarrow~~\infty $ \beq  S_{c0}^*=\frac{\Lambda}{R_0\mu}\left[1+\frac{\beta_1\sigma}{\beta_2(\gamma+\mu)}\right]\eeq
if $R_0-1>\frac{\beta_1\sigma}{\beta_2(\gamma+\mu)}$
\beqn I_{c0}^*&=&\frac{I_0}{R_0-1}\left[R_0-1-\frac{\beta_1\sigma}{\beta_2(\gamma+\mu)}\right], \nonumber \\
 E_{c0}^*&=&\frac{I_0}{R_0-1}\frac{\gamma+\mu}{\sigma}\left[R_0-1-\frac{\beta_1\sigma}{\beta_2(\gamma+\mu)}\right], \nonumber \\ R_{c0}^*&=&\frac{I_0}{R_0-1}\frac{\gamma}{\mu}\left[R_0-1-\frac{\beta_1\sigma}{\beta_2(\gamma+\mu)}\right], \nonumber \\ D_{c0}^*&=&\frac{I_0}{R_0-1}\frac{d\gamma}{\lambda}\left[R_0-1-\frac{\beta_1\sigma}{\beta_2(\gamma+\mu)}\right]\eeqn

For other values of $\kappa$ it can be shown that a single $0<I_{c0}^*$ exists if
\beq R_0-1>\frac{\beta_1\sigma}{\beta_2(\gamma+\mu)}\eeq
This can be proven graphically as shown in the Appendix.

\begin{figure}[!ht]
    \begin{center}
        \includegraphics[scale=0.5]{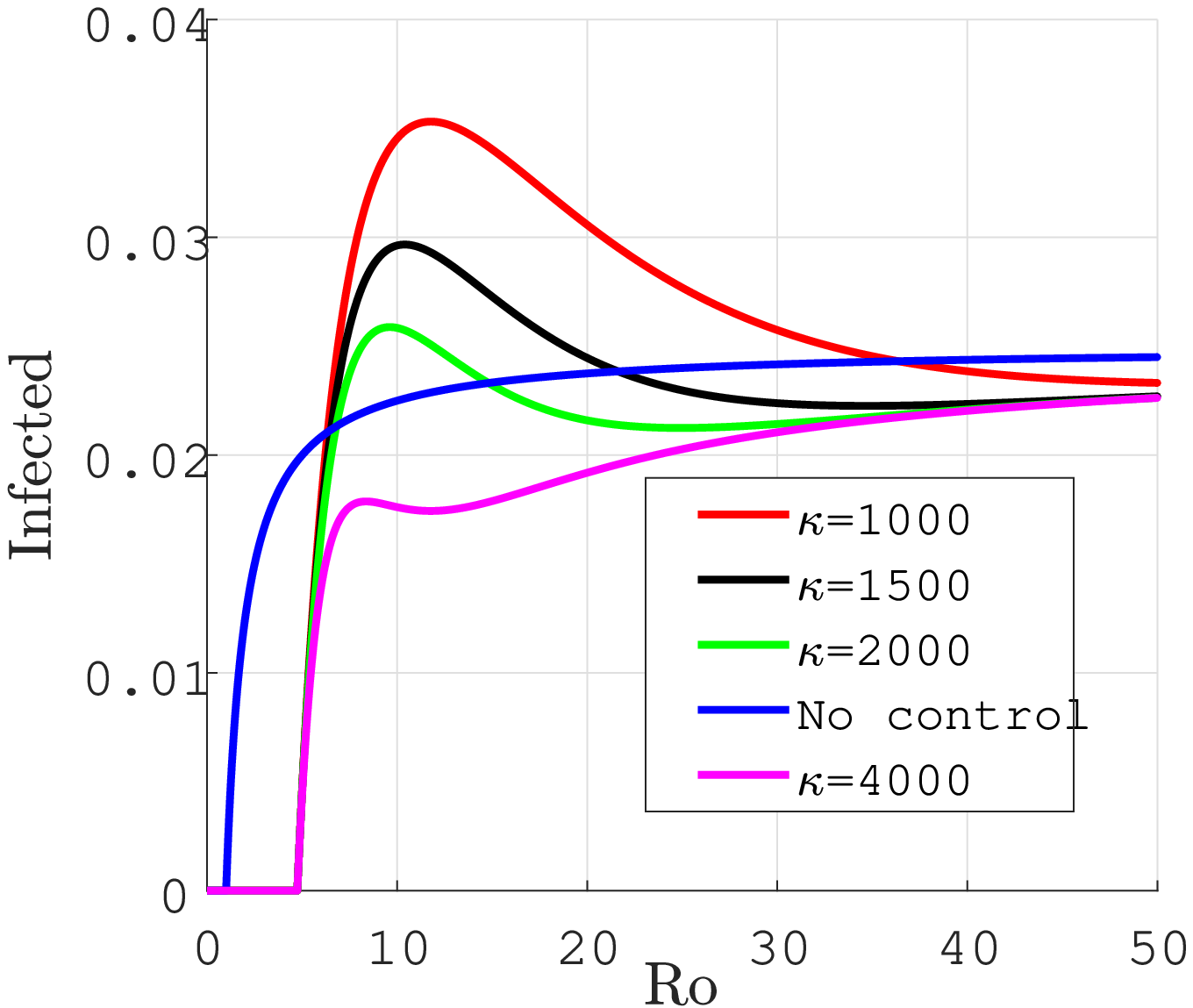}\\(a)
        \includegraphics[scale=0.5]{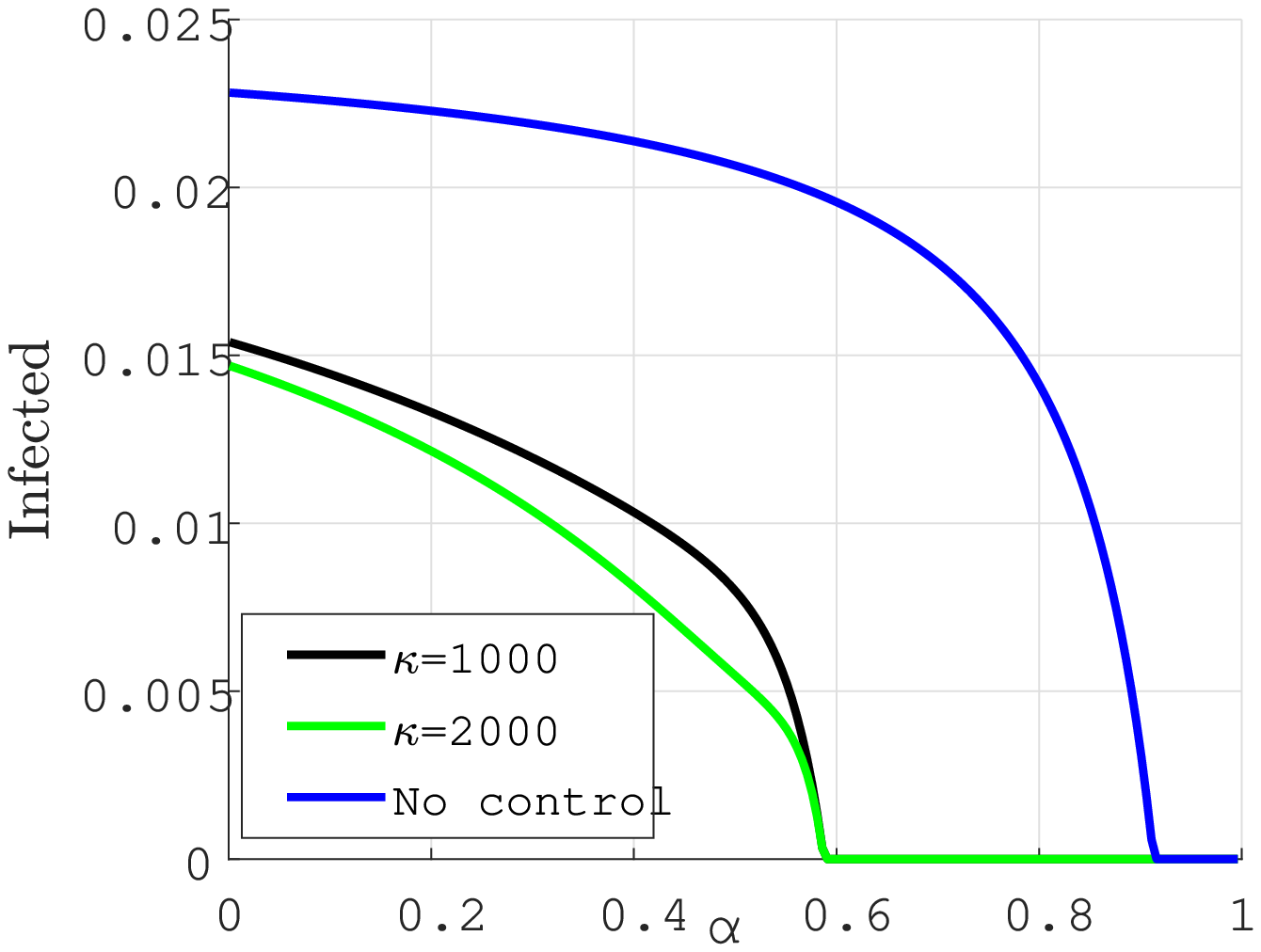}\\(b)
\caption{Effects of the intensity of the population response. (a): versus reproduction number. (b): versus government control}
\label{fig4}\end{center}
   \end{figure}

Figure~\ref{fig4} shows how the intensity of the population response could impact the spread of the disease. In fact, under this control, the number of infections is considerably reduced as shown in the figures. In Fig.~\ref{fig4}a, there is a jump in the number of infected for small $R_0$. This jump is significant for smaller value of $\kappa$ and is likely a manifestation of the nonlinearity in $\kappa$ in the expression of the infection function. Recalling that the endemic equilibrium used here was obtained for $\kappa\rightarrow\infty$, the results of Fig. \ref{fig4}a are only valid for larger values of $\kappa$. This nonlinearity is not visible in the presence of $\alpha$ as shown in Fig.~\ref{fig4}b.
\begin{figure}[!ht]
    \begin{center}
        \includegraphics[scale=0.5]{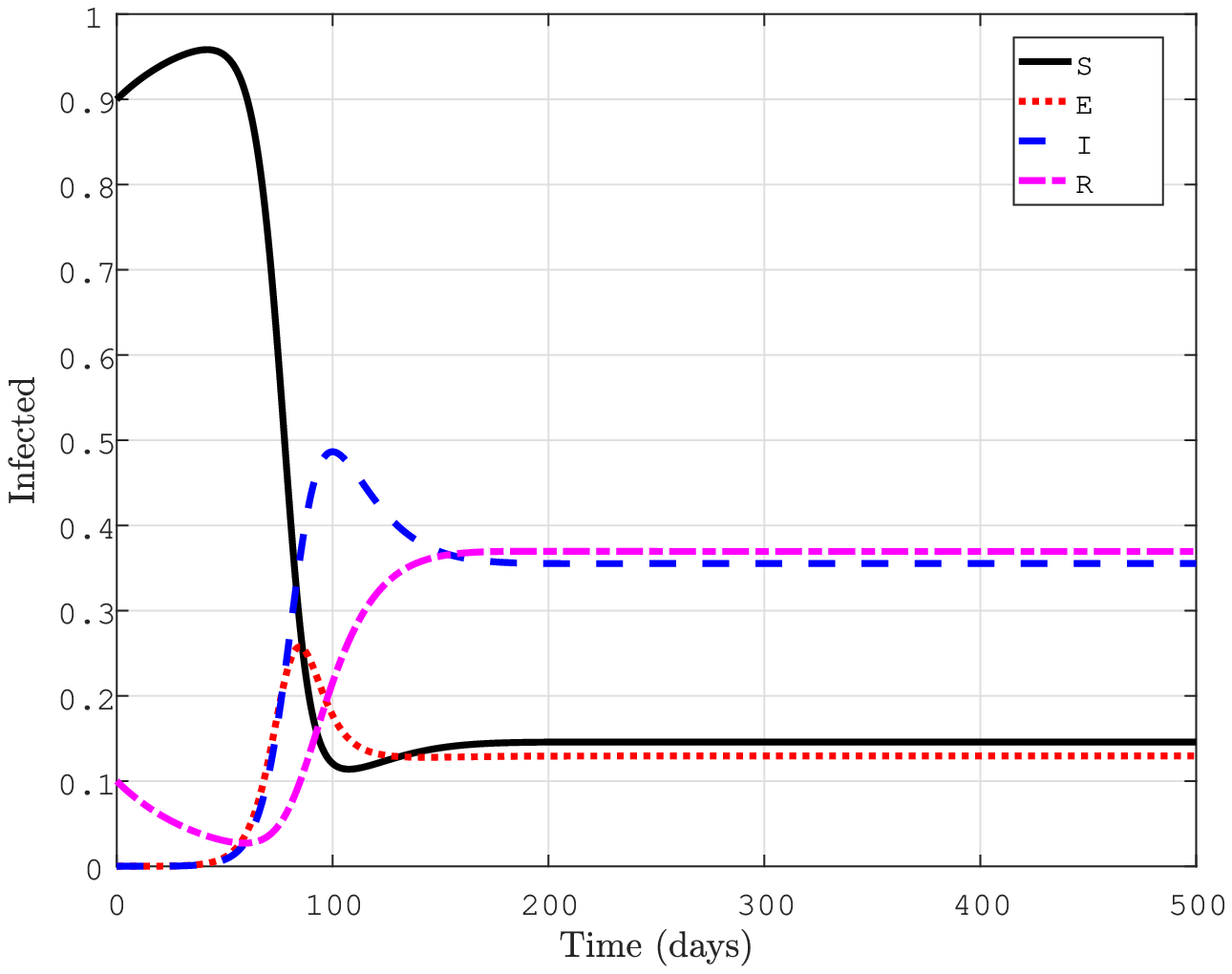}\\(a)
        \includegraphics[scale=0.5]{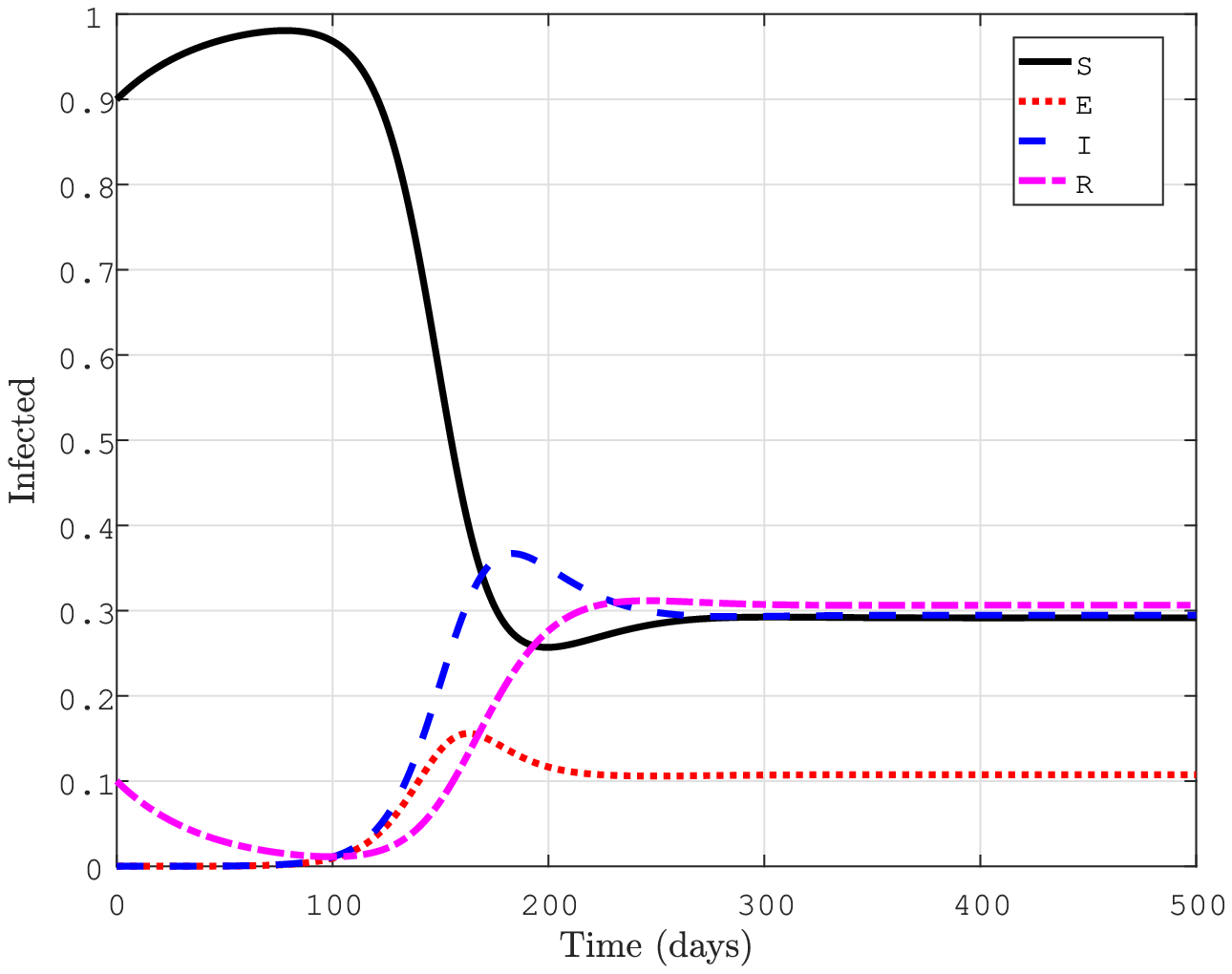}\\(b)
            \includegraphics[scale=0.5]{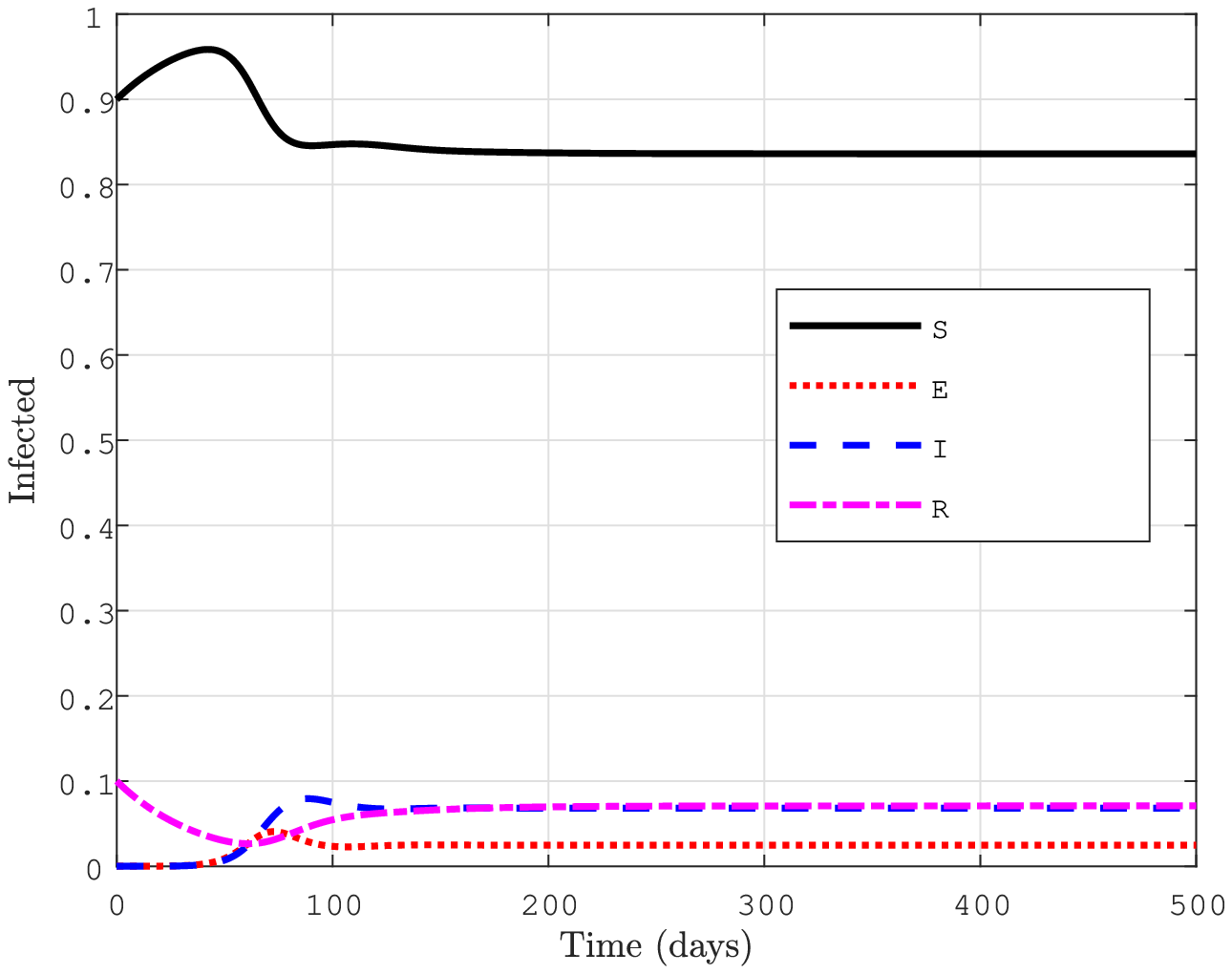}\\(a)
        \includegraphics[scale=0.5]{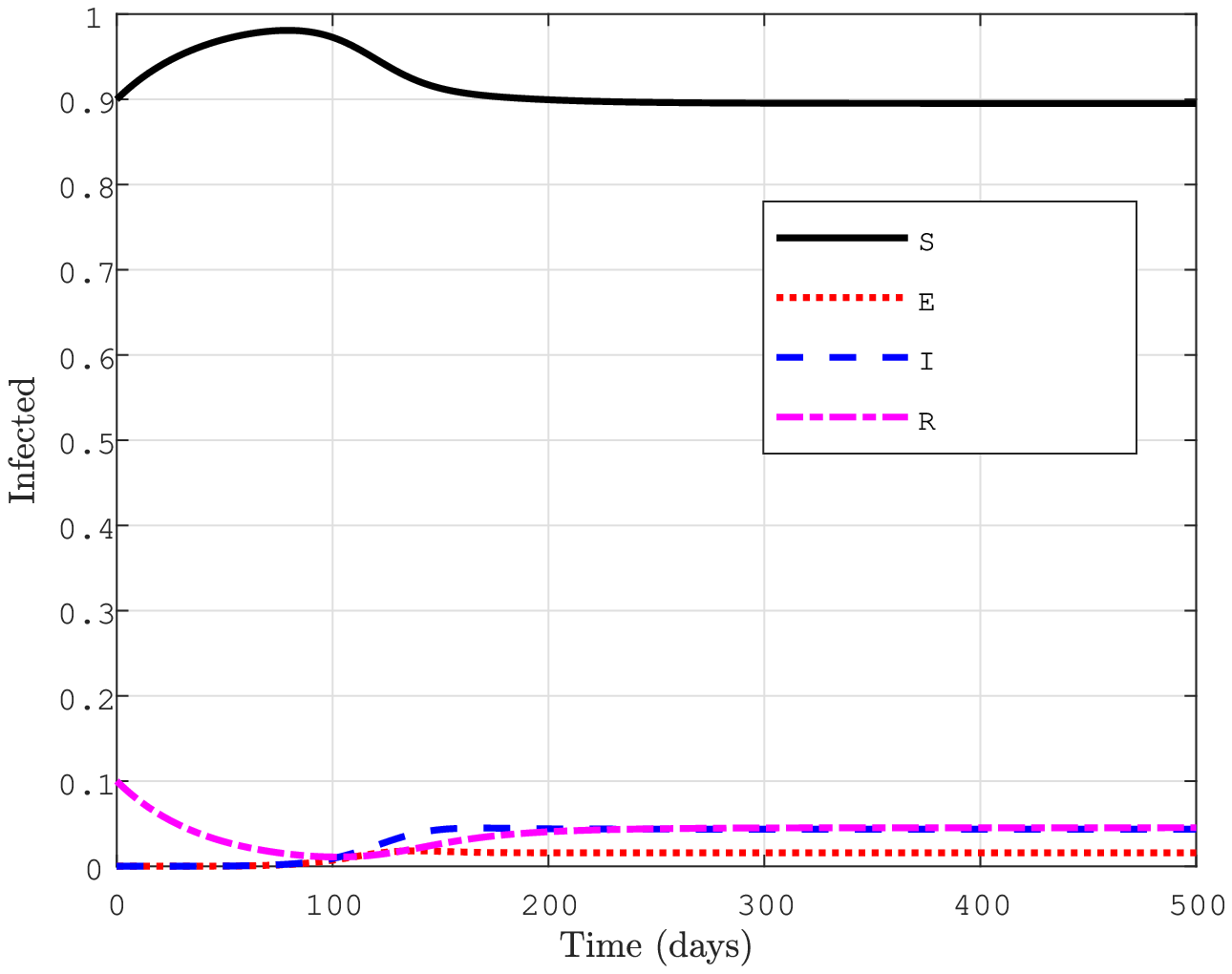}\\(b)
\caption{Response of the system under various scenarios. (a): Unlikely scenario of no control. (b): System with government control $\alpha$. (c): Effects of the population response $\kappa$. (d): Effects of combined control}
\label{fig7}\end{center}
   \end{figure}

Figure~\ref{fig7} shows an illustration of the system response for a na\"{i}ve scenario where there is no governmental action (Fig.~\ref{fig7}a), the effects of governmental action alone (Fig.~\ref{fig7}b), individual reaction alone (Fig.~\ref{fig7}c) and  combined action (Fig.~\ref{fig7}d). Simulation and  analytical derivation show that, carefully setting the parameters (in the specified range of vales) could effectively stop the spread of the disease under combined actions.

\section*{Appendix}

The existence of a unique endemic value of $I_{c0}^*$ for
\beq \label{stab} I_{c0}^*=\frac{I_0^*}{R_0-1}\left[R_0-\frac{\Lambda}{R_0\mu}\frac{\beta_1\sigma+ \beta_2(\gamma+\mu)}{\beta_1\sigma\left(1-\frac{d\gamma}{\lambda}I_{c0}\right)^{\kappa}+\beta_2(\gamma+\mu)}\right]
\eeq
can be shown graphically for all values of $\kappa$ by plotting the following graphs

\beq Z(X)=X,~~~Y(X)=\frac{I_0}{R_0-1}\left[1-\frac{1+N}{(1-mX)^{\kappa}}\right]\eeq
with $X\equiv I_{c0}^*$, $m=\frac{d\gamma}{\lambda}$ and $N=\frac{\beta_1\sigma}{\beta_2(\gamma+\mu)}$
The intersection point of $Z(X)$ and $Y(X)$ in the interval $[0,1]$ will exist if
\beq R_0-1>\frac{\beta_1\sigma}{\beta_2(\gamma+\mu)}\eeq
Figure~\ref{fig_fun} shows the plot of $Z(X)$ (in black line) and $Y(X)$ for several values of $\kappa$ from $\kappa=0$ (no perception) to realistic values of $\kappa$ ~\cite{Lin2020}. In all cases, the intersection of $Z(X)$ and $Y(X)$ is singular, thus there exists a unique $I_{c0}^*$ solution of Eq.~(\ref{stab}).
\begin{figure}[!ht]
    \begin{center}
        \includegraphics[scale=0.5]{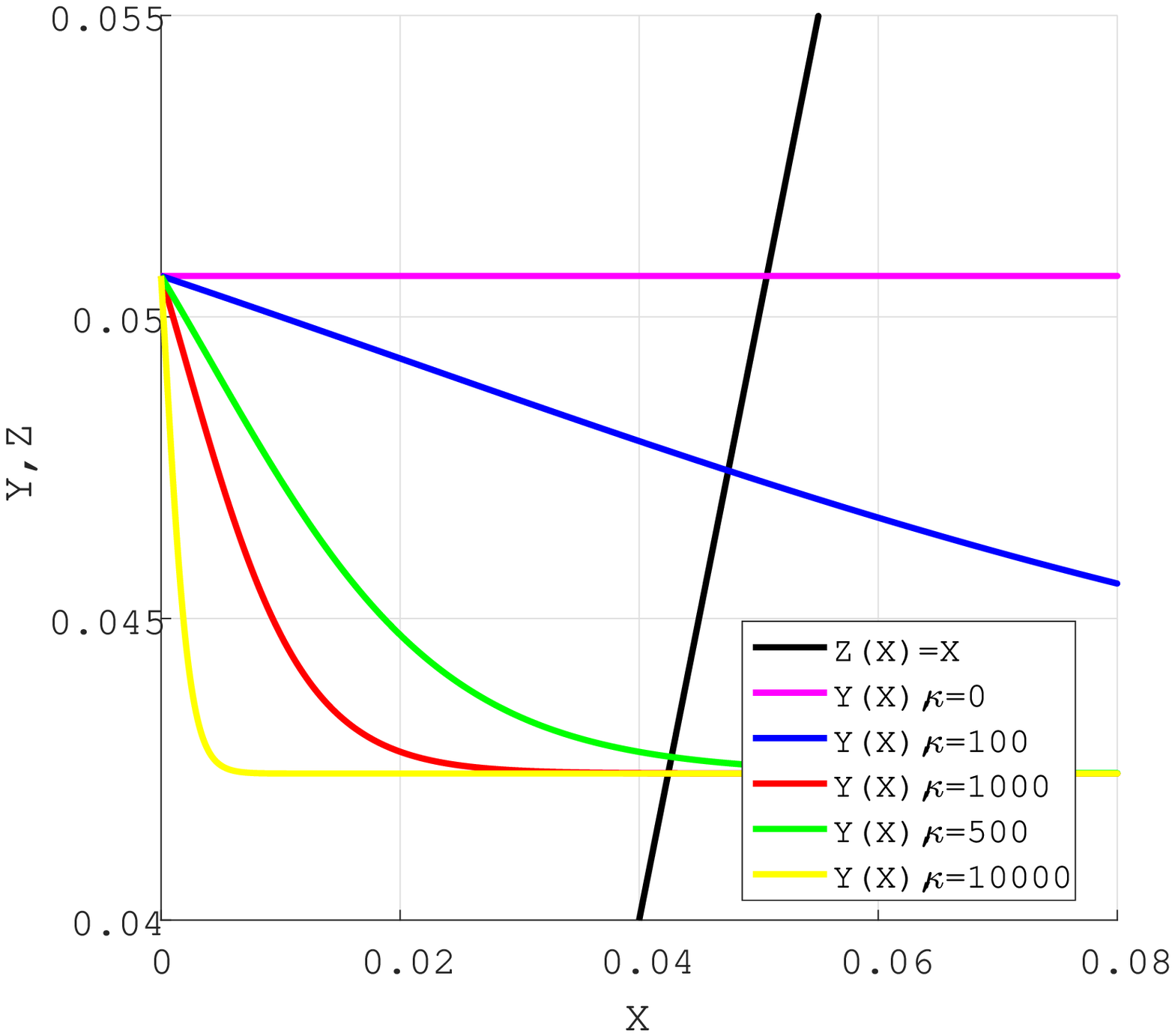}
\caption{Graphical illustration of the existence of a unique endemic equilibrium, with $\beta_1\beta_2=0.416$ and $\alpha=0$.}
\label{fig_fun}\end{center}
   \end{figure}


\section{Summary of Findings}

We summarize below the key findings of our analysis.
\begin{itemize}
\item The reproduction number, traditionally computed for SEIR model in terms of $\sigma,\Lambda,\mu$ and $\gamma$ has been expanded to include $\beta_1$, $\beta_2$, and the social and policy parameters, $\alpha$ and $\kappa$.  This expanded definition embeds social dynamics neatly into the epidemiological model and significantly expands insight into their interactions.
\item The parameter values for transmissibility ($\beta_1, \beta_2$), and hence, the reproduction number ($R_0$) went through significant reduction with the South Korean government response roughly 40 days after the first incidence.
\item Exposed people (as opposed to infected individuals) have a greater impact on the persistence of the disease.
\item The stronger the government action, the more the impact on disease transmission.
\item There is a minimum threshold value for government action ($\alpha_c$) for complete control of the disease. Our model predicts that numerous small, tentative steps would not be as effective as bolder and significant steps.
\item The intensity of the public response ($\kappa$) has significant impact on the reduction of number of infections.
\item The model predicts that for some values of the disease dynamics, the public perception $\kappa$ will have no effects. In this case only the governmental action could stop the spread of the disease.
\item The analysis predicts that a suitable combination of government response ($\alpha$) and public reaction ($\kappa$) would effectively stop pandemics such as COVID-19.
\end{itemize}


\section{Conclusion}

In this paper, we adapted and developed an SEIR model for the COVID-19 pandemic including different transmission rates for contacts with infected and exposed, and integrated parameters and variables to model government action and social reaction.  First, we used data from South Korea to perform a parametric analysis using the genetic algorithm and achieved a very good fit.  This provides sound validation for our model.   The resulting numerical analysis shows that the South Korean government action 40 days after the infection was first diagnosed had a significant influence on the spreading of the disease.

Next, we used more nuanced models for nonlinear dynamic analysis.  Equilibrium and stability analysis were performed revealing several areas of the parameter space where a stable endemic equilibrium can exist leading to persistent infections.  We considered three situations: (a) without control, (b) with government action, and (c) with combined effect of government action and public reaction.   Results show that it is possible to stop the spread of the disease (or to extinguish the endemic equilibrium) by proper choice of parameters that govern social and government behavior.

In this paper, by seamlessly integrating two important sociological (and arguably, political) parameters, i.e., public perception and government policy, we are able to show the fact that these factors can significantly affect the transmission rate and spread pattern of disease evolution. The conclusions would support an argument that stronger government actions and policies such as quarantine, wearing masks, social distancing, and improving public perception might be essential in combating the COVID-19 spread. Indeed, this is demonstrated in South Korea, which has arguably achieved tremendous success in combating COVID-19 unlike many other countries. Similar perspectives should be considered for further government policy regarding progressively reopening the economy and campuses. A potential future direction is to integrate more aspects including seasonal effects, which would likely lead to periodic responses.

Finally, as we write this paper, we note that the pandemic situation is still evolving with considerable uncertainty about the future. We believe that this paper demonstrates the importance of  {nonlinear dynamic analysis} to enhance our understanding of the natural world in which we the humans live and has profound implications for the way we handle it in the future.


\section{Acknowledgements}

CN \& FN gratefully acknowledge the financial support from US Office of Naval Research (Grant: N00014-19-1-2070) for basic research on adaptive modeling of nonlinear dynamic systems.  In particular, we appreciate the continuous encouragement from Capt. Lynn Petersen and are humbled by his recognition of the value of our research.

\section{Conflict of Interest}

The authors declare that they have no conflicts of interest.


\bibliographystyle{IEEEtran}
\bibliography{epidemiology}

\end{document}